\documentclass{article}
\usepackage[utf8]{inputenc}
\DeclareUnicodeCharacter{39D}{$\nu$}
\DeclareUnicodeCharacter{3BC}{$\mu$}
\DeclareUnicodeCharacter{39C}{M}
\DeclareUnicodeCharacter{2009}{\,}
\DeclareUnicodeCharacter{2212}{$-$}
\DeclareUnicodeCharacter{D7}{$\times$}
\usepackage{ragged2e}
\usepackage{textcomp}
\usepackage[style=nature]{biblatex}
\AtBeginDocument{
    \AtEveryBibitem{\clearfield{month}}
    \AtEveryBibitem{\clearfield{day}}
    \AtEveryBibitem{\clearfield{endmonth}}
    \AtEveryBibitem{\clearfield{endday}}
    \AtEveryBibitem{\clearfield{labelmonth}}
    \AtEveryBibitem{\clearfield{labelday}}
    \AtEveryBibitem{\clearlist{language}}
    \AtEveryBibitem{\clearfield{note}}
} 

\usepackage{amsmath}
\usepackage{authblk}
\usepackage{graphicx,xcolor}
\usepackage{upgreek}
\usepackage[T1]{fontenc}
\usepackage[labelfont={bf,sf},%
                labelsep=period,%
                figurename=Fig.,%
                singlelinecheck=off,%
                justification=justified,
                format=plain]{caption}
\usepackage[left=75pt,%
                right=75pt,%
                top=71pt,%
                bottom=85pt,%
                headheight=15pt,%
                headsep=10pt,%
                letterpaper,twoside]{geometry}%
\usepackage[super]{nth}

\title{20 years of developments in optical frequency comb technology and applications}
\author[\,]{Tara Fortier}
\author[\,]{Esther Baumann}
\affil[\,]{National Institute of Standards and Technology, Boulder, Colorado 80305, USA}
\affil[\,]{Department of Physics, University of Colorado, Boulder, Colorado 80309, USA}
\affil[\,]{Corresponding authosr: tara.fortier@nist.gov, esther.baumann@nist.gov}

\date{}

\begin{document}

\maketitle

% \tableofcontents

\section*{Abstract}

Optical frequency combs were developed nearly two decades ago to support the world’s most precise atomic clocks. Acting as precision optical synthesizers, frequency combs enable the precise transfer of phase and frequency information from a high-stability reference to hundreds of thousands of tones in the optical domain. This versatility, coupled with near-continuous spectroscopic coverage from the terahertz to the extreme ultra-violet, has enabled precision measurement capabilities in both fundamental and applied contexts. This review takes a tutorial approach to illustrate how 20 years of source development and technology has facilitated the journey of optical frequency combs from the lab into the field.

\section{Introduction} 

The optical frequency comb was originally developed to count the cycles from optical atomic clocks. Atoms make ideal frequency references because each atom is identical, and hence reproducible, with discrete and well-defined energy levels that are dominated by strong internal forces that naturally isolate them from external perturbations. Consequently, in 1967 the international standard unit of time, the SI second was redefined as 9,192,631,770 oscillations between two hyper-fine states in $^{133}$Cs \supercite{essenAtomicStandardFrequency1955}. While $^{133}$Cs microwave clocks provide an astounding 16 digits in frequency/time accuracy, clocks based on optical transitions in atoms are being explored as alternative references because higher transition frequencies (see Section \ref{sec:clocknet}) permit greater than a 100 times improvement in time/frequency resolution. Optical signals, however, pose a significant measurement challenge because light frequencies oscillate 100,000 times faster than state-of-the-art digital electronics. Prior to 2000, the simplest method to access an optical frequency was via knowledge of the speed of light and measurement of its wavelength, accessible with relatively poor precision of parts in $10^7$ using an optical wavemeter. For precision measurements, with resolutions better than that offered by wavelength standards, large-scale frequency chains were used to connect the microwave definition of the Hertz, provided by the $^{133}$Cs primary frequency reference near 9.2\,GHz, to the optical domain via a series of multiplied and phase-locked oscillators \supercite{hollbergMeasurementOpticalFrequencies2005}. The most complicated of these systems required up to 10 scientists, 20 different oscillators and 50 feedback loops to perform a single optical measurement \supercite{schnatzFirstPhaseCoherentFrequency1996}. Because of the complexity, frequency multiplication chains yielded one to two precision optical frequency measurements per year. In 2000, the realization of the optical frequency comb allowed for the replacement of these complex frequency chains with a single mode-locked laser, enabling vast simplification to precision optical measurement and rapid progress and development into optical atomic standards.

Optical frequency combs (OFCs) were developed by drawing on single-frequency laser stabilization techniques and applying them to mode-locked (pulsed) laser systems. The result was a system that could synthesize $10^5$ to $10^6$ harmonically-related optical modes from either an electronic or optical reference with a fidelity better than 1 part in $10^{18}$. More importantly, OFCs enabled the direct conversion of optical-to-microwave frequencies and vice versa, enabling the extraction of microwave timing signals from optical atomic clocks. Beyond their application to precision optical metrology, OFCs were quickly recognized for their versatility as high-fidelity optical frequency converters and as sources of precisely timed ultra-short pulses. More broadly, by taking advantage of the nonlinearities possible with the ultra-short pulses, OFCs enable synthesis over broad spectral regions including the near-infrared, the visible domain and as far as the extreme ultraviolet. Generation of difference frequencies within the optical spectrum also allows for high-fidelity frequency transfer to the mid-infrared, terahertz and microwave domains. Optical frequency combs quickly found application to a multitude of diverse optical, atomic, molecular and solid-state systems, including X-ray and attosecond pulse generation \supercite{kellerRecentDevelopmentsCompact2003}, coherent control in field dependent processes \supercite{baltuskaAttosecondControlElectronic2003, fortierCarrierEnvelopePhaseControlledQuantum2004}, molecular fingerprinting \supercite{diddamsMolecularFingerprintingResolved2007}, trace gas sensing in the oil and gas industry \supercite{aldenMethaneLeakDetection2017}, tests of fundamental physics with atomic clocks \supercite{rosenbandFrequencyRatioHg2008}, calibration of atomic spectrographs \supercite{murphyHighprecisionWavelengthCalibration2007}, precision time/frequency transfer over fiber and free-space \supercite{giorgettaOpticalTwowayTime2013}, arbitrary waveform measurements for optical communication \supercite{marin-palomoMicroresonatorbasedSolitonsMassively2017}, and precision ranging \supercite{minoshimaHighaccuracyMeasurement240m2000}. To support this broad application space, OFCs have seen rapid changes in laser development to enable coverage at different spectral regions, varying frequency resolutions, and to enable the development of systems that offer lower size, weight and power (SWAP) \supercite{delhayeOpticalFrequencyComb2007,manurkarFullySelfreferencedFrequency2018,yuGasPhaseMicroresonatorBasedComb2018, faistQuantumCascadeLaser2016}. 

The remarkable technical capabilities outlined above gained John "Jan" Hall and Theodor H\"ansch recpgnition by the Nobel Committee in 2005 for their life long contributions to the field of precision optical frequency metrology \supercite{hallNobelLectureDefining2006,hanschNobelLecturePassion2006}, as well as for their technical vision and expertise that resulted in the realization of the optical frequency comb \supercite{hallOpticalFrequencyMeasurement2000}. A quick search on Google Scholar for publications that contain the exact phrase, “optical frequency comb,” returns more than 14,000 publications on the topic in the last 20 years. In writing this review we hope to provide a broad historical overview of the origins of OFCs, and explain how they work and are applied in different contexts. More importantly, we hope to motivate the reader as to why they are such a powerful tool in the context of precision laboratory experiments, and explain how they are moving beyond the metrology laboratory and into the field.

\section{What is an optical frequency comb and how does it work}

The short answer is that an optical frequency comb is a phase stabilized mode-locked laser. While different generation methods have been developed since their first inception, mode-locked lasers provide the broadest capabilities for OFC operation and for its description. The utility of mode-locked lasers (MLL) within the context of optical metrology was recognized as early as the late 1980’s.  The optical pulses from modelocked lasers result from the coherent addition of 100's of thousands to millions of resonant longitudinal optical cavity modes, spanning up to 100\,nm in the optical domain. While the broad optical bandwidth is immediately attractive for spectroscopic applications, the mode-locked optical spectrum has unique properties that are beneficial for precision optical metrology: 1) all the optical modes are harmonically related (perfectly equidistant in frequency) and 2) all optical modes are phase coherent with one another (share a common phase evolution). 

The consequence of this is that the electric field, and consequently the phase and frequency dynamics of every optical mode in the laser spectrum is deterministic. This deterministic behavior allows for every mode in the optical spectrum to be described and controlled using only two characteristic microwave frequencies, $f_\mathrm{r}$ and $f_\mathrm{0}$. More importantly, the mode-locked laser spectrum enables direct and phase-coherent conversion between optical and microwave frequencies, their original claim to fame in precision optical metrology.

\begin{figure}[h!tb] 
    \centering
    \includegraphics[width=0.78\linewidth]{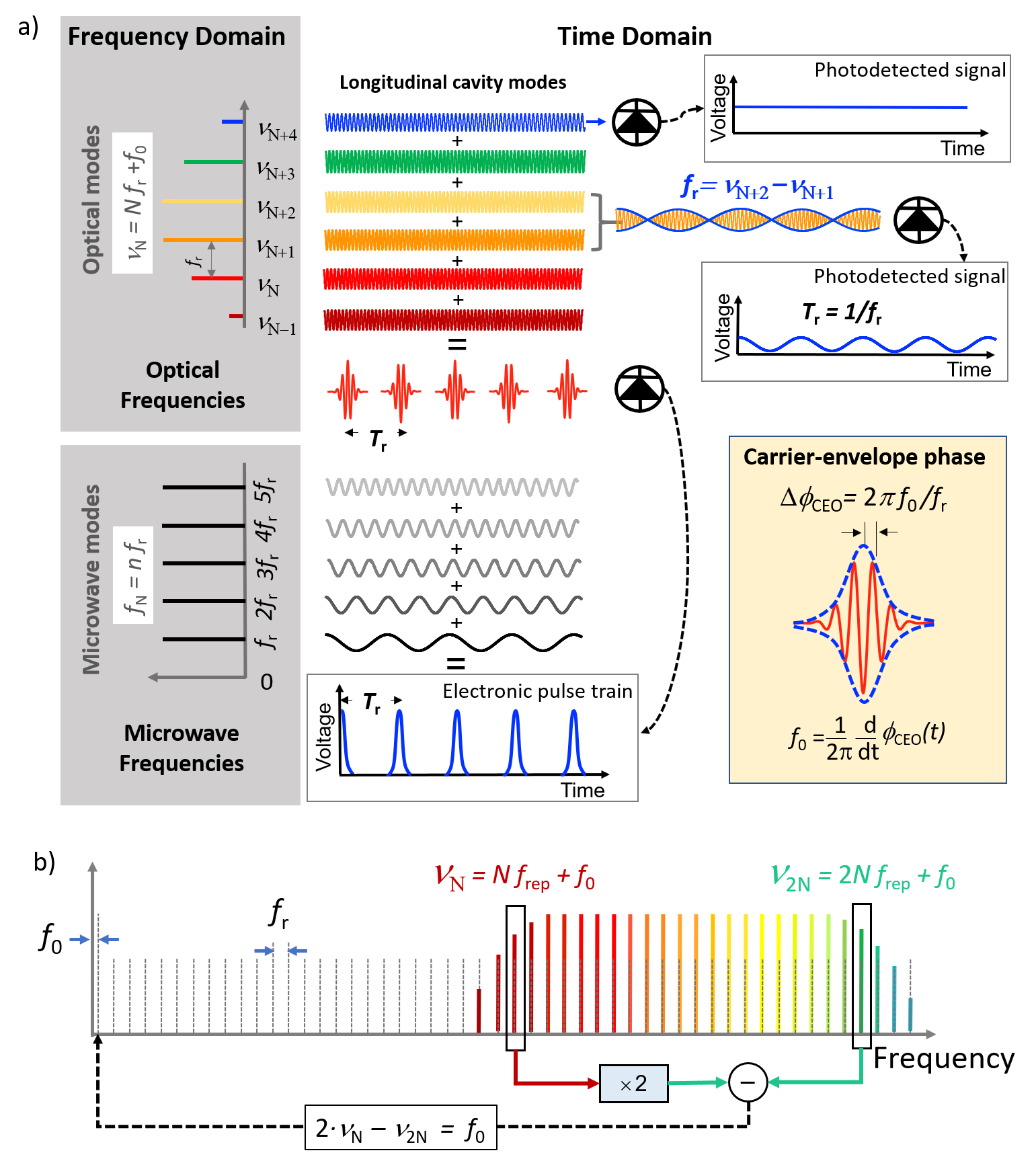}
    \caption{a) Time and frequency domain representation of an optical frequency comb. The optical output of a modelocked laser is a periodic train ot optical pulses with pulse period, $T_{\mathrm{r}}$, and pulse envelope $A(t)$. In the frequency domain, this pulse train can be expressed as a Fourier series of equidistant optical frequencies, with mode spacing, $f_{\mathrm{r}} = 1/T_{\mathrm{r}}$. The frequency of any optical mode, $\upnu_{\mathrm{N}}$, is characterized by only two degrees of freedom, $f_{\mathrm{r}}$ and $f_0$, such that $\nu_{\mathrm{N}} = N \cdot f_{\mathrm{r}} + f_0$. The laser repetition rate, $f_{\mathrm{r}}$, is accessed by detecting the amplitude modulation of the optical pulse train using a photodetector. This detection results in an electronic pulse train composed of coherently related microwave Fourier harmonics, $n \cdot f_{\mathrm{r}}$. Note that the optical spectrum contains $f_0$ information, whereas the microwave spectrum is only dependent on $f_{\mathrm{r}}$ because direct photodetection is not sensitive to the optical carrier. In the yellow shaded inset, we show the connection between $f_0$ and the carrier-envelope phase, $\phi_{\mathrm{CEO}}(t)$. The evolution in the pulse-to-pulse change in the carrier-envelope phase is given by $\Delta \phi_{\mathrm{CEO}} = 2\pi f_0/f_{\mathrm{r}}$. Notably, when $f_0 = 0$, every optical pulse has an identical carrier-envelope phase. The pulse envelope, $A(t)$, depicted by a blue dashed line is related by the periodic Fourier transform to the spectral envelope. b) Offset frequency detection via self-referencing. Frequency depiction of how nonlinear self-comparison can be used to detect $f_0 <f_{\mathrm{r}}$, which is manifest as a common offset that shifts the optical comb spectrum from perfect integer harmonic of $f_{\mathrm{r}}$.}
    \label{fig:comb_timedomain}
\end{figure}

\subsection{The comb equation}

The most succinct frequency description of the capabilities in the previous section is contained in the comb equation, which describes the deterministic relationship between the harmonic modes of the modelocked optical spectrum. As seen in Fig. \ref{fig:comb_timedomain}, it is the regular frequency spacing of the modes in the optical spectrum that inspired the analogy to a comb, although the analogy to a frequency ruler better describes the OFCs measurement capability. To understand the origins of the comb equation, we will quickly explore the relatively simple mathematics that describe the optical field output from a modelocked laser.

The optical field of the laser pulse train can be described by a carrier frequency, $\nu_{\mathrm{c}} = \omega_{\mathrm{c}}/(2\pi)$, that is modulated by a periodic pulse envelope, $A(t)$. Typically, the time between optical pulses range between 1 to 10\,ns. Due to the pulse periodicity, the optical field can also be described as a periodic Fourier series of optical modes, $\nu_{\mathrm{N}} = \omega_{\mathrm{N}}/(2\pi)$, with Fourier amplitude components, $A_{\mathrm{N}}$, such that

\begin{equation}
    E(t) 
    = A(t)e^{i \omega_{\mathrm{c}} t} = \sum_{N=N_i}^{N_f}A_{\mathrm{N}} e^{iN\times\omega_{\mathrm{N}} t}.
    \label{equation:electric_field}
\end{equation}

\noindent Because $\nu_{\mathrm{c}}$ is not necessarily an exact multiple of the mode spacing, $f_{\mathrm{r}}$, the individual Fourier frequencies are shifted from integer multiples of $f_{\mathrm{r}}$ by a common offset, $f_0 \leq f_{\mathrm{r}}$, such that 

\begin{equation}
    \nu_{\mathrm{N}} = N \cdot f_{\mathrm{r}} + f_0, 
    \label{eq:opt_mode}
\end{equation}

\noindent where $N$ is an integer mode number between 100,000 and 1,000,000, that multiplies $f_{\mathrm{r}}$ from the microwave domain to the optical domain.

Equation \ref{eq:opt_mode} is referred to as the comb equation. What the comb equation states is that while an OFC consists of up to a million optical modes, spanning hundreds of terahertz in the optical domain, only two degrees of freedom: 1) the repetition rate, $f_{\mathrm{r}}$ and the 2) laser offset frequency, $f_0$, are needed to define the frequency of each individual optical mode, $\nu_{\mathrm{N}}$.

\vspace{3mm}
\textbf{The repetition rate ($f_{\mathrm{r}}$).}
The microwave mode that ties the spectrum together harmonically is the laser repetition rate, $f_{\mathrm{r}}$, which is the inverse of the pulse-to-pulse timing, $T_{\mathrm{r}}$. Pulses exit the laser cavity once per round trip such that the pulse repetition period, $T_{\mathrm{r}} = 2L/v_{\mathrm{g}}$, where $v_{\mathrm{g}}$ is the pulse group velocity in the laser cavity, is defined and controlled via actuation of the laser cavity length, $L$. Changes in $f_{\mathrm{r}}$ result in an accordion-like expansion and contraction of the frequency modes.

\vspace{3mm}
\textbf{The offset frequency ($f_0$).} 
For the coherent additional of longitudinal laser modes for pulse formation requires that every mode is perfectly equidistant in frequency and shares a common phase. This unlikely condition is enforced by nonlinearity in the cavity that underlies pulse formation, which shifts the phase of every individual optical mode to enable modelocking. The resulting coherence between the laser optical modes is manifest as the common and additive frequency offset, $f_0$, which acts to translate all the laser modes simultaneously. Because this offset frequency is a measure of coherence it also relates to time-changes of the optical carrier phase relative to the pulse envelope, $\phi_{\mathrm{CEO}}(t)$, 

\begin{equation}
    f_0=(1/2\pi)\cdot \mathrm{d}\phi_{\mathrm{CEO}}/\mathrm{d}t
    \label{eq:f0}
\end{equation}

\noindent that result due to dispersion induced phase- and group- velocity differences.

\vspace{3mm}
In the simplest terms, $f_{\mathrm{r}}$ controls the pulse-to-pulse timing, and hence the periodicity of the pulse train, permits coarse frequency control of the OFC spectrum, and connects the optical and microwave domains via $N f\mathrm{r}$. The offset frequency, $f_0$, controls the carrier-phase of the pulse train, and enables fine optical frequency tuning. The detection of the laser offset frequency $f_0$ is the key for allowing precise frequency determination of the comb modes. Without knowledge of the offset frequency, a single optical mode can only be known to $\pm f_{\mathrm{r}}$. On an optical frequency, this represents an error of parts in 10$^{6}$ to 10$^{5}$) depending on the mode spacing. Additionally, control over the pulse-to-pulse carrier-envelope phase evolution is central to enabling "quantum control" in field-sensitive atomic and molecular systems \supercite{baltuskaAttosecondControlElectronic2003}.

Full frequency stabilization of the comb is achieved using negative feedback to the laser cavity length and intra-cavity dispersion to physically control $f_{\mathrm{r}}$ and $f_0$. Ensuring good mechanical stability and that some care is taken in mechanical and electronic engineering of the stabilization loops, the above methods can enable control of the average cavity length at resolutions below a femtometer, the diameter of the proton. Applications with the highest stability requirements, or ones that require long-term accuracy and averaging, generally require stabilization of both $f_{\mathrm{r}}$ and $f_0$. As will be discussed later in the text, out-of-lab applications that use OFCs to measure Doppler broadened molecular linewidths, or applications that do not benefit from perfectly controlled environments, can use OFCs with lower stability and accuracy (see section \ref{sec:outoflab}).

\subsection{The offset frequency and measurement of the comb parameters}
\label{sec:meascombparameters}

While it is impossible to count optical frequencies directly, optical difference frequencies are easily accessible as long as they fall within the bandwidth limit of precision frequency counters near 10\,GHz. As a simple example, consider two optical carriers close in frequency, $\nu_1$ and $\nu_2$, that will interfere to produce an optical carrier with an amplitude modulation at the difference frequency, $\Delta f = \nu_1 - \nu_2$, (see Fig. \ref{fig:comb_timedomain} a)). When this signal is incident on a photodetector, the detector produces a voltage proportional to the amplitude modulation. This signal is often referred to in the literature as a heterodyne optical beat frequency. This technique of difference frequency measurement is at the heart of nearly all measurement techniques with optical OFCs, and enables access to its characteristic frequencies, $f_0$ and $f_{\mathrm{r}}$. 

\vspace{3mm}
\textbf{Detection of $f_{\mathrm{r}}$.}
The pulse train that is output from a mode-locked laser is essentially a massively amplitude modulated optical carrier because pulse formation results from the interference of all $10^5$ to $10^6$ modes of the OFC spectrum This amplitude modulation is the extreme case of the example of two mode beating in the previous paragraph. The more modes that contribute, the shorter the optical pulses and the stronger the amplitude modulation. Because mode-locked lasers used as OFCs typically have optical cavity lengths that vary between 30\,cm to 3\,m, $f_{\mathrm{r}}$ is an easily accessible microwave frequency between 1\,GHz and 100\,MHz, respectively. Direct photodetection of the optical pulses results in an electronic signal that only follows the amplitude modulation of the pulse train. As seen in Fig. \ref{fig:comb_timedomain} a), the frequency decomposition, or Fourier Transform, of the resulting electronic pulses yield harmonics of $f_{\mathrm{r}}$, but yield no information about $f_0$. Said otherwise, direct optical heterodyne between two optical modes of the comb only yields information about $f_{\mathrm{r}}$ because $f_0$ is common to each mode, or $\nu_{\mathrm{N}}-\nu_{\mathrm{M}} = N\cdot f_{\mathrm{r}} +f_0 - (M \cdot f_{\mathrm{r}} + f_0) = (N-M)\cdot f_{\mathrm{r}}$.

\vspace{3mm}
\textbf{Detection of $f_0$.} As explained previously, the fact that $f_0$ is related to the phase of the optical carrier makes it extremely difficult to access directly. In 1999 a method was proposed to produce a heterodyne beat at $f_0$ \supercite{telleCarrierenvelopeOffsetPhase1999} by nonlinear self-referencing between the extremes of the optical comb spectrum [see Fig. \ref{fig:comb_timedomain} b)]. The simplest manifestation of this technique is obtained by frequency doubling light from a comb mode on the low end of the optical comb spectrum and interfering it with fundamental light at twice the frequency such that
\begin{equation}
    f_0=2\cdot \nu_{\mathrm{N}} - \nu_{\mathrm{2N}}=2\cdot (N f_{\mathrm{r}}+f_0)-(2N\cdot f_{\mathrm{r}} +f_0).
    \label{eq:f0b}
\end{equation}

While mathematically simple, realization of this method requires that the OFC spectrum span an optical octave of bandwidth. This was problematic because optical spectra output from the broadest mode-locked lasers was < 100\,nm, significantly less than an optical octave. For comparison, an optical octave of bandwidth from a Ti:Sapphire laser center at 800\,nm constitutes a hefty 500\,nm, or 1000\,nm of bandwidth for an Er:fiber laser centered at 1550\,nm. 

\vspace{3mm}
\textbf{Continuum generation}
While high-energy ultra-short laser pulses were being explored in the 1990s for few-cycle pulse generation and spectroscopic applications, it was developments in highly-engineered low-dispersion optical fiber that enabled continuum generation at lower pulse energies \supercite{dudleyTenYearsNonlinear2009a}. These small core (1 to 3\,$\upmu$m) silica fibers balanced material dispersion with a waveguide dispersion enabled via fiber tapering \supercite{birksSupercontinuumGenerationTapered2000}, or via air cladding supported by glass webbing \supercite{rankaOpticalPropertiesHighdelta2000}. When ultra-short pulses from a Ti:Sapphire laser were launched into these fibers, the combination of small cross section, and low dispersion allowed for high-pulse intensities to be maintained over interaction lengths of several centimeters up to several meters. The result is coherent white light continuum generation, that, quoting directly from the text in Birks \textit{et al.} \supercite{birksSupercontinuumGenerationTapered2000}, "has the brightness of a laser with the bandwidth of a light bulb." It was only a matter of months after these first demonstrations that optical frequency combs with Ti:Sapphire lasers were fully realized \supercite{apolonskiControllingPhaseEvolution2000, telleCarrierenvelopeOffsetPhase1999, holzwarthOpticalFrequencySynthesizer2000}. 

The ability to directly convert optical frequencies to the microwave domain and vice versa resulted in a rapid advance in precision metrology capabilities. Within the first four years of their realization nearly everything that could be proposed with OFCs was demonstrated. This included carrier-envelope phase control \supercite{jonesCarrierEnvelopePhaseControl2000}, the first all-optical atomic clocks \supercite{diddamsOpticalClockBased2001}, absolute optical frequency measurements and the measurement of optical atomic frequency ratios \supercite{udemOpticalFrequencyMetrology2002}, searches for variation of fundamental constants \supercite{bizeTestingStabilityFundamental2003}, precision distance measurement \supercite{minoshimaHighaccuracyMeasurement240m2000}, coherent bandwidth extension and single cycle pulse synthesis \supercite{sheltonPhaseCoherentOpticalPulse2001}, coherent and direct microscopy \supercite{potmaHighsensitivityCoherentAntiStokes2002}, direct molecular spectroscopy \supercite{keilmannTimedomainMidinfraredFrequencycomb2004}, the development of molecular frequency references \supercite{yeMolecularIodineClock2001}, and optical synthesis of precision electronic signals \supercite{diddamsDesignControlFemtosecond2003}. It was absolute madness!

\section{Emergence of comb sources, frequency generation and new architectures}

\subsection{Evolution of solid-state and fiber-based mode-locked combs}
\label{sec:modlockOFC}

Once the stabilization techniques were understood, any mode-locked laser system that had sufficiently high pulse energy ($\sim$ 1\,nJ) for broadening, could be converted to an OFC. As a result, highly non-linear fibers for continuum generation were studied extensively to enable extension to different wavelengths. Advances in fiber technology along with the desire for more energy efficient OFCs yielded diode-pumped solid state lasers systems near 1\,$\upmu$m based on Cr:LiSAF, Yb:CALGO, Yb:KGW, Er:Yb:glass and Yb:KYW \supercite{schiltCarrierEnvelopeOffsetStabilized2015}, as well as diode-pumped fiber systems emitting light in the telecommunication band around 1550\,nm. The latter fiber lasers, built with off-the-shelf all-fiber components, allowed for even more compact, energy efficient and robust systems. Subsequently these Er:fiber-OFCs have seen the most commercial success and are the most commonly used OFCs system to date \supercite{hartlIntegratedSelfreferencedFrequencycomb2005, baumannHighperformanceVibrationimmuneFiberlaser2009, sinclairInvitedArticleCompact2015, hanselAllPolarizationmaintainingFiber2017, xiaRecentDevelopmentsFiberbased2016, manurkarFullySelfreferencedFrequency2018}. Other notable fiber lasers include the high power 1\,$\upmu$m Yb:fiber \supercite{ruehlAdvancesYbFiber2012}, which when combined with high power Ytterbium amplifiers are ideal candidates for high-harmonic generation in the XUV for direct comb spectroscopy \supercite{cingozDirectFrequencyComb2012} and the realization of 2\,$\upmu$m Thulium doped fiber-OFCs \supercite{jiangFullyStabilizedSelfreferenced2011}. In more recent years, bandwidth extension to the mid IR has seen research into non-silica based fiber lasers such as that based on Er$^{3+}$:fluoride \supercite{duvalFemtosecondFiberLasers2015}.

The use of the above solid-state and fiber-based systems, combined with continuum generation in non-silica based nonlinear fibers currently provides near-continuous and coherent spectroscopic coverage from 400\,nm to $\sim 4\,\upmu$m \supercite{priceMidIRSupercontinuumGeneration2007,granzowMidinfraredSupercontinuumGeneration2013}. In many ways, frequency generation with short-pulsed laser systems enables the only means for broad spectroscopic coverage at some wavelengths with high brightness. This is particularly true in the mid-IR to the terahertz \supercite{schliesserMidinfraredFrequencyCombs2012,weichmanBroadbandMolecularSpectroscopy2019} and the ultraviolet (UV) to the extreme ultraviolet (XUV) \supercite{gohleFrequencyCombExtreme2005,pupezaHighpowerSubtwocycleMidinfrared2015}. The more extreme demonstration of bandwidth extension has been to wavelengths as long as 27\,$\upmu$m \supercite{kowligyInfraredElectricField2019} using a combination of difference frequency generation (DFG) and/or optical parametric oscillation (OPO) \supercite{schliesserMidinfraredFrequencyCombs2012,gambettaMilliwattlevelFrequencyCombs2013,iwakuniGenerationFrequencyComb2016,ycasHighcoherenceMidinfraredDualcomb2018,muravievMassivelyParallelSensing2018}. Difference frequency generation relies on phase matching in standard nonlinear crystals to down-convert two photons of higher energy, a pump $\nu_1$ and signal $\nu_2$, to an idler mid-IR photon via $\nu_3=\nu_1 - \nu_2$. Optical parametric oscillators can perform either DFG or sum frequency generation (SFG, $\nu_3=\nu_1 + \nu_2$) in a resonant optical cavity, which permits highly efficient and extremely versatile frequency conversion. Below 400\,nm, researchers borrowed techniques from pulsed table-top X-ray sources to generate high optical harmonics by focusing cavity-enhanced optical pulses into a jet of noble gas for the production of UV and XUV frequencies. The highest achievable coherent frequency generation resulted from the 91$^{\mathrm{st}}$ harmonic at 11\,nm from a 60\,W Yb:fiber laser focused into a gas jet of Argon \supercite{pupezaHighpowerSubtwocycleMidinfrared2015}.

\begin{figure}[h!tb] 
    \centering
    \includegraphics[width=0.95\linewidth]{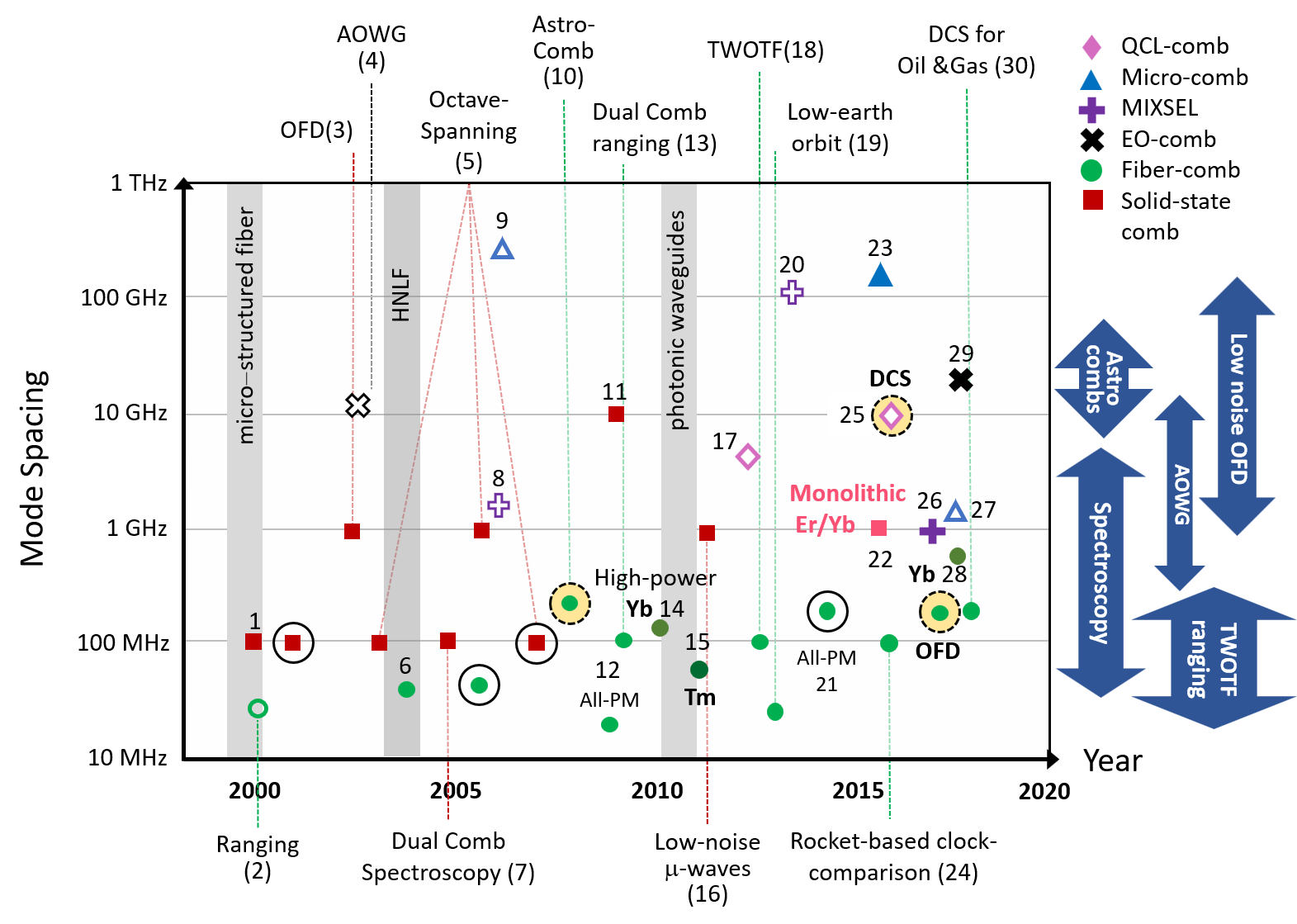}
    \caption{Overview of the development of optical frequency comb sources as function of year. The left axis indicates the mode-spacing of the various sources. To the right of the graph we indicate what mode-spacing range is most suitable for various applications. Milestones in source development, as well as some notable applications, beyond and including some of those listed in Section \ref{sec:meascombparameters}, are indicated at the bottom and top of the graph.  Filled markers indicate systems that have accessed $f_0$, while empty markers have not. Sources that have become commercial products are circled with a solid outline and comb-based products are circled with a dashed outline and filled in yellow. AOWG - arbitrary optical waveform generation, OFD - optical frequency division, TWOTFT - two-way optical time and frequency transfer, DCS - dual-comb spectroscopy.
    List of references:
    1:  \supercite{apolonskiControllingPhaseEvolution2000,telleCarrierenvelopeOffsetPhase1999,holzwarthOpticalFrequencySynthesizer2000}
    2:  \supercite{minoshimaHighaccuracyMeasurement240m2000}
    3:  \supercite{diddamsDesignControlFemtosecond2003}
    4:  \supercite{fujiwaraOpticalCarrierSupply2003}
    5: \supercite{fortierOctavespanningTiSapphire2006,bartelsGenerationBroadbandContinuum2002,matosDirectFrequencyComb2004}
    6: \supercite{washburnPhaselockedErbiumfiberlaserbasedFrequency2004}
    7: \supercite{schliesserFrequencycombInfraredSpectrometer2005, keilmannTimedomainMidinfraredFrequencycomb2004}
    8: \supercite{maasVerticalIntegrationUltrafast2007}
    9: \supercite{delhayeOpticalFrequencyComb2007}
    10: \supercite{steinmetzLaserFrequencyCombs2008}
    11: \supercite{bartels10GHzSelfReferencedOptical2009a}   
    12: \supercite{baumannHighperformanceVibrationimmuneFiberlaser2009}
    13: \supercite{coddingtonRapidPreciseAbsolute2009}
    14: \supercite{ruehl80120Fs2010}
    15: \supercite{jiangFullyStabilizedSelfreferenced2011}
    16: \supercite{fortierGenerationUltrastableMicrowaves2011}
    17: \supercite{hugiMidinfraredFrequencyComb2012}
    18: \supercite{giorgettaOpticalTwowayTime2013}
    19:  \supercite{leeTestingFemtosecondPulse2014}
    20: \supercite{mangoldPulseRepetitionRate2014}
    21: \supercite{hanselAllPolarizationmaintainingFiber2017}
    22: \supercite{shojiUltralownoiseMonolithicModelocked2016}
    23: \supercite{braschSelfreferencedPhotonicChip2017}
    24:\supercite{leziusSpaceborneFrequencyComb2016}
    25: \supercite{klockeSingleShotSubmicrosecondMidinfrared2018}
    26: \supercite{jornodCarrierenvelopeOffsetFrequency2017}
    27: \supercite{suhGigahertzrepetitionrateSolitonMicrocombs2018}
    28: \supercite{maLownoise750MHzSpaced2018}
    29: \supercite{carlsonUltrafastElectroopticLight2018}
    30: \supercite{aldenBootstrapInversionTechnique2018}
    }
    \label{fig:sources_year}
\end{figure}

\subsubsection{Current state-of-the-art in mode-locked laser sources}

After their first demonstrations, much effort in mode-locked laser OFCs design was placed on satisfying the dual goals of higher performance and lower SWAP. To keep pace with improvements in optical atomic clock development, common-mode measurement architectures and higher-bandwidth actuators have enabled long-term fidelity in optical frequency synthesis with MLLs \supercite{nicolodiSpectralPurityTransfer2014,leopardiSinglebranchErFiber2017} at parts in $10^{20}$. In the past two decades MLL sources have evolved from 80\,MHz, 2-m long Ti:Sapphire laser systems to gigahertz repetition rate, directly octave spanning Ti:Sapphire OFCs, to highly environmentally stable, all-polarization maintaining, fully-fiberized compact Er:fiber OFCs \supercite{sinclairInvitedArticleCompact2015}, and finally to monolithic, high-performance, sub 100\,fs Er/Yb:glass lasers that can fit in the palm of one hand \supercite{shojiUltralownoiseMonolithicModelocked2016}, see Fig. \ref{fig:sources_year}. 

\subsection{Compact and chip-scale sources}
\label{sec:compactOFC}
The past 15 years have also seen the development of even more compact and lower SWAP systems based on microresonators and semiconductor systems. Also described in this section are electro-optic frequency combs, which are currently the only source with a vastly tunable repetition rate, but that face similar challenges as semiconductor and microresonator systems described below. The compact size of these systems yield great excitement about the possibility for chip-scale and photonically integrated OFC sources \supercite{torres-companyOpticalFrequencyComb2014,carlsonUltrafastElectroopticLight2018,gaetaPhotonicchipbasedFrequencyCombs2019,faistQuantumCascadeLaser2016,gaafarModelockedSemiconductorDisk2016}.
 
\subsubsection{Semiconductor lasers}

Different semiconductor laser platforms have been investigated as OFC sources including quantum cascade lasers (QCL) \supercite{hugiMidinfraredFrequencyComb2012,sterczewskiTerahertzHyperspectralImaging2019} and mode-locked integrated external-cavity surface-emitting lasers (MIXSELs) \supercite{linkDualcombModelockedLaser2015,gaafarModelockedSemiconductorDisk2016}. MIXSELs are vertical emitting semiconductor lasers integrated with semiconductor saturable absorber mirrors, which help induce modelocking. When optically pumped, MIXSEs can produce sub-100\,fs pulses and greater than 1\,W of optical power. In addition, the integrated semiconductor platform is a potential candidate for mass production with substantially reduced fabrication costs and high-efficiency, and can be engineered to operate from 800\,nm to the near-IR. Whereas the operating wavelength of a traditional semiconductor laser is determined by the bandgap of the material, QCLs rely on sandwiched quantum-well heterostructures that behave as engineered bandgap materials. As a result, the quantum cascade laser offers a versatile system based on four-wave mixing \supercite{faistQuantumCascadeLaser2016} for the generation of mid-IR to terahertz radiation with variable mode-spacing from 5 to 50\,GHz. While QCL-combs do not produce optical pulses, which results in serious challenges to nonlinear broadening, stability, and regularity in mode spacing, they currently offer the only OFC platform with direct electrical pumping.

\subsubsection{Micro-resonator systems}

Optical frequency combs based on micro-resonators, or micro-combs, differ significantly in operation from MLLs because they are not lasers, but low-loss, optical resonators. The first of these systems developed as OFCs were based on suspended silica micro-torroids, \supercite{delhayeOpticalFrequencyComb2007} and machined and hand-polished crystalline CaF$_2$ micro-rods \supercite{savchenkovTunableOpticalFrequency2008}. Micro-resonator architectures have since expanded to more easily integrated and lithographically engineered and patterned waveguides based on a multitude of materials, whose various properties are summarized in Ref.\supercite{gaetaPhotonicchipbasedFrequencyCombs2019}. Micro-resonators act as build up cavities that enable high-nonlinearity over long storage times, or equivalently long-interaction lengths, in very much the same way as do nonlinear fibers. Via degenerate- and non-degenerate four-wave mixing, a resonantly coupled single-frequency pump source is converted to a comb of optical frequencies. Using these systems, coherent octave-spanning bandwidths have been observed for micro-resonators with $\simeq 200$\,GHz native mode spacing \supercite{braschSelfreferencedPhotonicChip2017}. However, the optical bandwidth narrows significantly for larger resonator diameters that permit more accessible mode spacing. While micro-combs enable chip-scale comb generation, they do not directly yield optical pulses. Because pulse formation is crucial for coherent comb formation, much of the early micro-resonator work was aimed at understanding the temporal dynamics of stable optical soliton production, which is now regularly realized via systematic and careful control of the pump
laser detuning \supercite{gaetaPhotonicchipbasedFrequencyCombs2019,suhMicroresonatorSolitonDualcomb2016}.

\subsubsection{Electro-optic comb generators}

The equally spaced optical modes of frequency comb generators based on a phase-modulated single-frequency laser were used in precision optical metrology prior to 2000 \supercite{hallOpticalFrequencyMeasurement2000} to bridge and measure smaller frequency gaps (<10’s of terahertz) between the last multiplication stage of the frequency chain to an unknown transition of interest (see section \ref{sec:optfrqmeas}) \supercite{kourogiWidespanOpticalFrequency1993}. Because of their simple optical architecture and the possibility for multi-gigahertz mode spacing derived directly from an electronic synthesizer, these sources have been revisited in recent years primarily in the context of arbitrary optical waveform generation \supercite{cundiffOpticalArbitraryWaveform2010} and high bit-rate optical communication and microwave photonics \supercite{torres-companyOpticalFrequencyComb2014}. Perhaps the most versatile feature of electro-optic combs (EO-combs) is the fact that they are the only OFC source that offers wide-band and agile tuning of the mode spacing. An additional benefit of these systems is the availability of high-speed electro-optic modulators that operate at pump wavelengths spanning 780\,nm to 2\,$\upmu$m. More recently, electro-optic combs (EO-combs) have yielded access to $f_0$ for full stabilization \supercite{carlsonUltrafastElectroopticLight2018}. Full stabilization has lead to the impressive applications of an EO-comb to the calibration of an astronomical spectrograph \supercite{metcalfStellarSpectroscopyNearinfrared2019}, which requires frequency stability at 1 part in $10^{11}$ and an ultra-wide and extremely flat optical spectrum (see Section \ref{sec:astroOFC}).

\subsubsection{Super-continuum generation below 200\,pJ}
\label{sec:waveguides}

To date, nearly all high repetition rate and compact frequency comb systems \supercite{delhayePhasecoherentMicrowavetoopticalLink2016, jornodCarrierenvelopeOffsetFrequency2017, carlsonUltrafastElectroopticLight2018} required high-power optical amplification to enable fiber-based continuum generation for detection of $f_0$.  This is because the combination of high repetition rate, low output power, narrow optical bandwidth, as well as no modelocking mechanism in the case of EO- or QCL-combs, results in pulse energies that are more than 100 times lower than those possible with mode-locked lasers. Continuum generation in lithographically patterned photonic waveguides and nanowires (as well as extremely small core chalcogenide fibers \supercite{granzowMidinfraredSupercontinuumGeneration2013}) have recently helped to mitigate these difficulties. These photonic waveguides have been to compact combs, what nonlinear fiber was to MLLs. The use of extremely small cross-sections ($< 1\,\upmu\mathrm{m} \times 1\,\upmu\mathrm{m}$), waveguide dispersion engineering, broad transparency windows, and a higher non-linear index as compared to silica fiber, have permitted super-continuum generation at much lower pulse energies (< 200\,pJ)  \supercite{halirUltrabroadbandSupercontinuumGeneration2012}. Much like the developments that took place with nonlinear fibers, the past decade has seen the remarkable engineering of photonic waveguides for continuum generation from the visible to the mid-IR \supercite{kuykenMidinfraredTelecombandSupercontinuum2011,halirUltrabroadbandSupercontinuumGeneration2012,heckUltralowLossWaveguide2014,porcelTwooctaveSpanningSupercontinuum2017,yoonohCoherentUltravioletNearinfrared2017,guoMidinfraredFrequencyComb2018,hicksteinUltrabroadbandSupercontinuumGeneration2017,gaetaPhotonicchipbasedFrequencyCombs2019}. Because patterned waveguides are lithographically produced, they offer more versatility than optical fibers in terms of spectral shaping, but on the downside suffer significantly higher insertion and propagation losses.

\subsubsection{Current state-of-the-art in compact and chip-scale sources}

Although gains have been made in the continuum generation techniques described above, semiconductor and micro-resonators systems have yet to achieve their full potential as compact systems for precision optical comb generation. Optically pumped MIXSELs have been fully stabilized, with low residual noise, using extended cavities ($f_{\mathrm{r}} < 2\,\mathrm{GHz}$) and external amplification \supercite{jornodCarrierenvelopeOffsetFrequency2017}, but at the compromise of SWAP and system complexity. Of the compact sources, QCL-based OFCs currently offer the lowest SWAP due to the fact that they are electrically pumped. Despite the fact that QCLs currently do not offer access to $f_0$ for precision operation, their low-SWAP and access from the mid-IR to terahertz have allowed these sources to emerge as a commercially available platforms for dual-comb spectrometers \supercite{klockeSingleShotSubmicrosecondMidinfrared2018}.

In micro-resonators, stable soliton formation has been understood and can now be regularly controlled, and broadening to an optical octave has been achieved to enable coherent detection of $f_0$ \supercite{delhayePhasecoherentMicrowavetoopticalLink2016,braschSelfreferencedPhotonicChip2017}. However, soliton formation in micro-resonator systems requires significant detuning of the pump away from the resonator modes, resulting in an optical efficiency of less than 3\,$\%$ \supercite{gaetaPhotonicchipbasedFrequencyCombs2019}. Consequently, the low optical output power necessitates the use of high-power external optical amplifiers and/or “helper” lasers for frequency doubling to enable $f_0$ detection. These auxiliary optical components currently limit the photonic integration of micro-combs and increases their SWAP. Despite these challenges, microresonators have been applied to many of the experiments that traditionally employ modelocked lasers, but with degradation in performance \supercite{spencerOpticalfrequencySynthesizerUsing2018,duttOnchipDualcombSource2018, trochaUltrafastOpticalRanging2018, marin-palomoMicroresonatorbasedSolitonsMassively2017}. Currently the largest obstacle to achieving low-drift and narrow optical linewidths in microresonators is thermorefractive noise, which results because microscopic systems naturally exhibit high sensitivity to both temperature and pressure. An additional blue-detuned auxiliary laser can be used to correct thermally-induced changes in the material index of refraction by nearly 10\,dB \supercite{drakeThermalDecoherenceLaser2019}, but once again at the expense of power consumption and complexity.

Although compact sources based on micro-resonator and semiconductor systems still face challenges to full optical integration, to date these platforms offer the only possible architectures for chip-based and integrated comb systems. The future integration of compact OFCs with CMOS compatible photonic waveguides based on, for example lithium niobate, silicon or silicon nitride, diode lasers and miniature optical clocks might one day enable sub-watt systems for both optical \supercite{spencerOpticalfrequencySynthesizerUsing2018} and microwave synthesis, enabling cost-efficient production for dissemination of optical frequency comb sources and products to larger commercial markets.

\section{Optical frequency comb applications}
\label{sec:app}

In the following sections we explain how optical frequency combs have impacted various applications and their evolution. For simplicity, we group applications into two categories based on application stability requirement. The first section focuses on the application of OFCs to high precision frequency synthesis and measurement, which require the most stringent requirements to enable the comparison and dissemination of signals for atomic clocks, Section \ref{sec:clockscomp}. Section \ref{sec:outoflab} encompasses applications that have begun to move from the laboratory toward more commercial applications. These latter applications typically require more robust and less environmentally sensitive sources, and can tolerate lower stability requirements because of their use for direct molecular spectroscopy and distance measurements in less controlled environments. While the above experiments do not encompass the full application space enabled by OFCs, we limit the discussion to these topics due to the length constraints of the manuscript.

 \begin{figure}[h!tb] 
    \centering
    \includegraphics[width=0.68\linewidth]{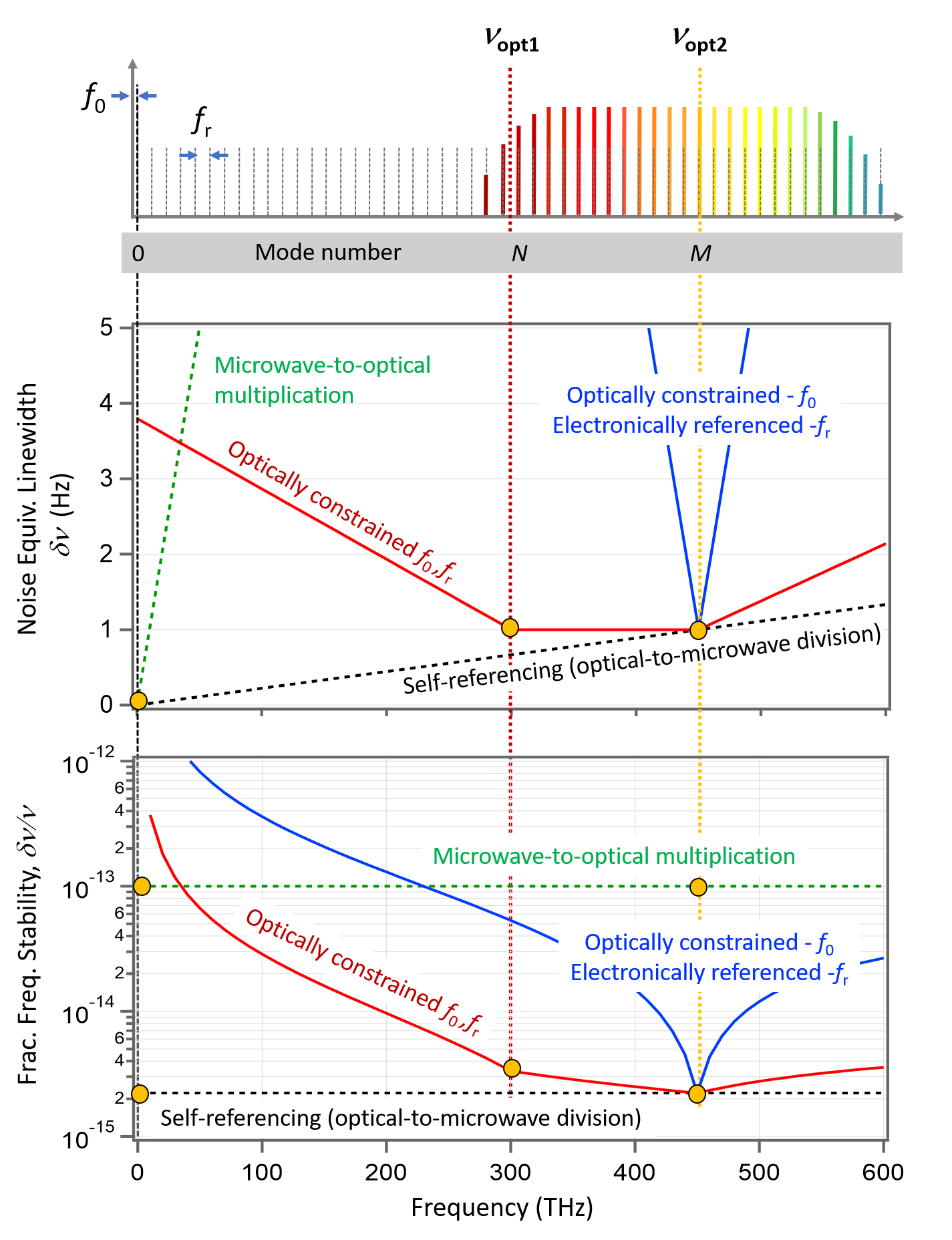}
    \caption{Optical frequency comb performance as a function of referencing scheme. This performance is constrained to systems where the noise does not overwhelm the carrier. The metric for performance is indicated as both the linewidth, $\delta\nu$ of the modes as a function of frequency as well as the fractional frequency stability, $\delta\nu/\nu$, where $\nu$ is the carrier frequency of the mode. Constraint of the two degrees of freedom, namely $f_0$ and $f_{\mathrm{r}}$ is obtained via either direct referencing to a microwave reference, or by indirect referencing of an optical mode to an optical reference. For clarity, the referencing points are indicated as filled yellow circles. In the red trace, the two optical comb modes are referenced against two optical sources with 1\,Hz linewidths. In the blue trace, $f_{\mathrm{r}}$ is compared directly to a microwave reference, while $f_0$ is constrained via comparison to an optical reference. Both the blue and red referencing schemes can be used for combs that do not have direct access to $f_0$. For the green and the black dashed trace, $f_0$ is compared directly to a microwave reference near DC (the linewidth of the microwave reference appears to be zero on the linear scale because a 100 MHz synthesized signal with  10$^{-13}$ fractional stability, has a noise equivalent linewidth of < 10\,$\upmu$Hz ). For the green trace, $f_{\mathrm{r}}$ is referenced directly in the microwave domain, whereas for the black dashed trace, $f_{\mathrm{r}}$ is constrained via an optical mode to a 1\,Hz optical reference.}
    \label{fig:referencing}
\end{figure}

\subsection{Comb referencing, stabilization and performance}
\label{sec:OFCstab}
Because the optical spectrum from a mode-locked laser is not stable on its own, precise knowledge and control of both comb parameters is required to harness the laser's full potential for precision metrology. The measurement of any frequency, optical or microwave, requires comparison against a second frequency reference. Here we explore how the chosen references and the frequency (microwave or optical) at which a comb is referenced impacts the stability and noise of its modes. Knowledge and control of the comb modes can be achieved by comparing $f_{\mathrm{r}}$ and $f_0$ directly to microwave references, by constraining the two degrees of freedom in the optical domain to optical references, or via a combination of the two methods.

Aside from full phase stabilization, passive optical and electronic phase noise removal from the comb parameters can be used for frequency measurement using of an OFC. These methods include difference frequency generation (DFG) for the creation of optical combs with vanishing $f_0$\supercite{puppeCharacterizationDFGComb2016}, and synchronous electronic "mix-out" of $f_0$ and $f_{\mathrm{r}}$ using what is called the transfer oscillator method\supercite{telleKerrlensModelockedLasers2002}. Full frequency stabilization uses negative feedback to the laser cavity length and intra-cavity dispersion to actively and physically control the values of $f_{\mathrm{r}}$ and $f_0$ via comparison to chosen frequency references. These are most typically achieved using piezoelectric-actuators for length control and modulation of the laser pump power and intra-cavity gain, which actuates on the laser intra-cavity dispersion. In contrast, the transfer oscillator method loosely controls the drift of $f_0$ and $f_{\mathrm{r}}$ so that their electronic signals can be precisely tracked to correct their noise excursion across the microwave and optical modes of the comb \supercite{telleKerrlensModelockedLasers2002,deschenesOpticalReferencingTechnique2010,nicolodiSpectralPurityTransfer2014}. 

\textbf{Passive $f_0$ removal via DFG}. Difference frequency generation can be used to create a frequency comb with vanishing $f_0$ information. The result is an "offset-free" comb with optical modes that are described by multiples of $f_{\mathrm{r}}$ only, whereby $\nu_k = \nu_N-\nu_M = (N-M)\cdot f_{\mathrm{r}}+f_0-f_0$ = $N' \cdot f_{\mathrm{r}}$. This passive removal of $f_0$ eliminates the need for its measurement and referencing (while OPOs can synthesize idler frequencies based on DFG, $f_0$ cancellation is not necessarily a given because the signal is created via nonlinear conversion from the pump in the OPO). Creating optical combs via DFG requires $\nu_{\mathrm{M}}$ and $\nu_{\mathrm{N}}$ to be separated by large difference frequencies. For example, a MLL centered at 1550\,nm can produce an "offset-free" comb at 3\,$\upmu$m, but requires an octave of bandwidth ($\sim$ 1000\,nm) \supercite{puppeCharacterizationDFGComb2016}. Additionally, the low-efficiency and phase matching requirements of the nonlinear crystal results in low-power and narrow bandwidth DFG combs. As a result, the DFG comb often requires both optical amplification and nonlinear broadening for power and bandwidth recovery.

\vspace{3mm}
Figure \ref{fig:referencing} shows how the OFC noise performance varies as a function of different referencing schemes. For readers that are more familiar with noise analysis, the arguments here are constrained to systems where the noise does not significantly overwhelm the optical carrier \supercite{newburyLownoiseFiberlaserFrequency2007}. The stabilization schemes depicted by the blue and red traces can be used to constrain $f_0$, for a system where its detection is not accessible. The performance in black and green traces are only possible with combs whereby $f_0$ is directly accessible, or "offset-free" comb. In the red trace in Fig. \ref{fig:referencing} we see that the noise equivalent linewidth, $\delta\nu$ \supercite{didomenicoSimpleApproachRelation2010} of the comb modes is constrained between the two lock points, but diverges outside the lock points with a slope $\delta\nu_K$ given by the noise of the references (added in quadrature), divided by their spacing, 

\begin{equation}
  \delta\nu_{\mathrm{K}} = \sqrt{\delta\nu_{\mathrm{ref1}}^2+\delta\nu_{\mathrm{ref2}}^2}/|(M-N)|
  \label{Eq:mode_noise}
\end{equation}

Equation \ref{Eq:mode_noise} indicates that the further apart the lock points, the better the frequency leverage to constrain all the comb modes. As a result, the self-reference stabilization method (black trace) exhibits the highest performance by having lock points separated by 100’s of terahertz, with $f_{\mathrm{r}}$ constrained via stabilization of a mode in the optical domain, and by direct stabilization of $f_0$ to a microwave reference near 0.

The self-referenced locking scheme also enables optical frequency division, whereby both the frequency and noise of the optical reference is divided down to the microwave domain \supercite{hollbergOpticalFrequencyWavelength2005,mcferranLownoiseSynthesisMicrowave2005,fortierGenerationUltrastableMicrowaves2011,xiePhotonicMicrowaveSignals2017}. In this scheme, the divided optical signal is detected via the repetition rate, which carries the noise of the optical reference at mode $M$ divided by $M^2$, $\delta f_{\mathrm{r}} = S_{\phi,\mathrm{ref} }/M^2$. To put this division into context, an optical reference at 300\,THz divided down and detected on a repetition frequency of 10\,GHz, would yield a reduction of the optical phase noise to the microwave domain by 90\,dB (= 10 $\cdot$Log(300\,THz / 10\,GHz)$^2$, or 9 orders of magnitude. The green trace in Fig. \ref{fig:referencing}, which also exhibits a linear dependence between noise and mode number, uses frequency multiplication to transfer a microwave reference to the optical domain. In both schemes the phase noise power spectral density (PSD) of mode number $K$, $S_{\phi,\mathrm{K} }$ is equal to the phase noise PSD of the reference, $S_{\phi,\mathrm{ref} }$ (at mode $M$), scaled by the square of the ratio of the mode numbers, or

\begin{equation}
    S_{\phi,\mathrm{K} }\sim  S_{\phi,\mathrm{ref} }(K^2/M^2).
    \label{eq:locking1}
\end{equation}

Because both the frequency and noise scale equally in multiplication and division, use of either the green or black traces for stabilization yields preservation of the fractional frequency stability of the reference in its transfer to all comb modes, or

\begin{equation}
    \sigma = \delta \nu_{\mathrm{ref}}/\nu_{\mathrm{ref}}=\delta \nu_{\mathrm{K}}/\nu_{\mathrm{K}}.
    \label{eq:locking2}
\end{equation}

\subsection{Combs for atomic clock comparisons}
\label{sec:clockscomp}

The full characterization of the optical spectrum of the mode-locked laser in the year 2000 allowed for frequency multiplier chains to be replaced by a single mode-locked laser \supercite{hallOpticalFrequencyMeasurement2000,diddamsDirectLinkMicrowave2000,stengerPhasecoherentFrequencyMeasurement2001,udemAbsoluteOpticalFrequency2000}. As will be discussed in more detail in the following section, this increased measurement capability was marked by the rapid characterization of unknown optical clock transition frequencies. The large bandwidth of optical frequency combs also allowed for the relative measurement of different species of developing optical atomic clocks \supercite{udemOpticalFrequencyMetrology2002,ludlowOpticalAtomicClocks2015}, enabling frequency comparisons below the limit imposed by the $^{133}$Cs primary reference.

\subsubsection{Clock comparisons}
\label{sec:optfrqmeas}

\begin{figure}[h!tb] 
    \centering
    \includegraphics[width=0.8\linewidth]{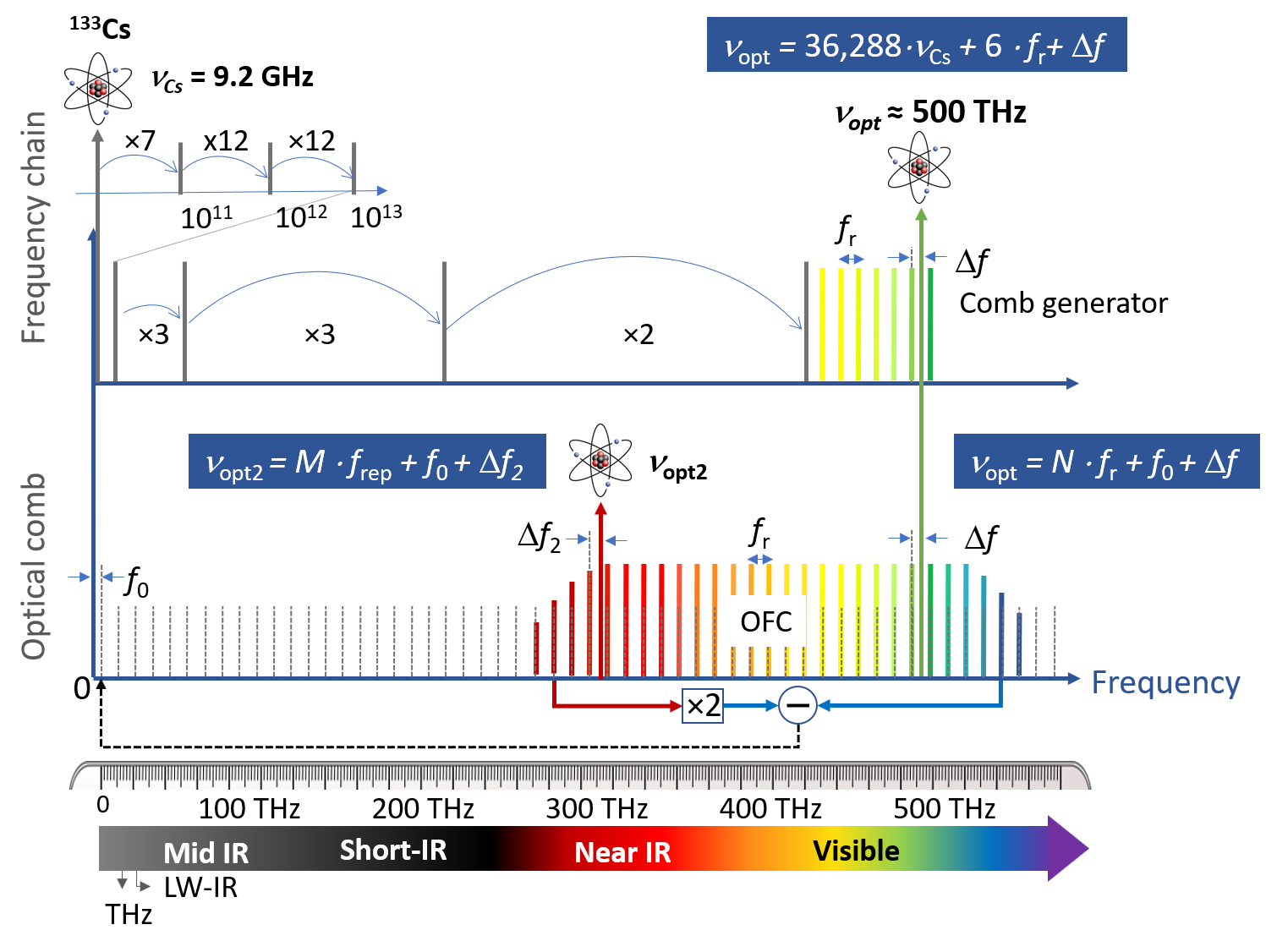}
    \caption{Schematic showing how frequency chains and frequency combs measure an unknown optical transition frequency. Before the year 2000, optical comb generators were used to bridge smaller frequency gaps by stabilizing the mode spacing, $f_{\mathrm{r}}$, directly to a microwave reference. Because the offset frequency of these systems could not be measured directly, to constrain $f_0$, a mode of the comb generator was locked to the optical output from the last multiplication stage of the frequency chain. In contrast, an optical frequency comb (OFC), which enables access to $f_0$, the lower frequency end of the optical frequency comb is pinned in the microwave instead of in the optical domain. This entirely eliminates the need for a frequency multiplier chain and permits frequency division of the optical reference to the microwave domain. We also depict how the offset frequency from an OFC is detected, and how an OFC can allow for the relative comparison of two optical clocks via knowledge of their offset frequencies, $\Delta f$ and $\Delta f_2$, with respect to their nearest OFC modes, $N$ and $M$, respectively.}
    \label{fig:comb_freqdomain}
\end{figure}

\textbf{Absolute frequency measurement}
As mentioned briefly in the introduction, optical atomic clocks were developed because their higher transition frequencies enabled better frequency/time resolution than their microwave counterparts. Generally, as described in Eq. \ref{eq:locking2}, the resolution of an oscillator is defined as the uncertainty with which the frequency can be defined, $\delta\nu$, scaled by its center frequency, $\nu$. More specifically, the frequency uncertainty of atomic clocks is limited by the clock transition sensitivity to external environmental fields, as well as how well these fields can be controlled and/or measured. Because the measurement and control of the latter physical parameters (temperature, E- and B-fields, optical power, and background pressure) is roughly the same for microwave and optical clocks, the benefit of the high transition frequency in optical clocks is two-fold; one can achieve both higher resolution with less stringent environmental control. To put this into perspective, to achieve 10$^{-16}$ fractional stability on a 500\,THz optical clock requires control of the optical transition frequency at the 50\,mHz level. Conversely, the same fractional stability on a 10\,GHz signal requires control of a microwave transition frequency at the 1\,$\upmu$Hz level. The upshot is that optical atomic clocks still have significant room for improvement, both in terms of SWAP and performance, whereas the performance of microwave clocks has plateaued.

As seen in Fig. \ref{fig:comb_freqdomain}, frequency combs and comb generators measure frequency in much the same way that rulers measure distance. They work by generating optical frequencies on an equidistant grid, with a known grid spacing. An unknown optical frequency, $\nu_{\mathrm{opt}}$, is determined by measuring the difference frequency, $\Delta f$, between it and the nearest comb mode. Prior to the year 2000, frequency comb generators were used to bridge smaller gaps in frequency. This enabled a significant simplification to multiplier chains that required additional oscillators to frequency shift various multiplication stages such that the last stage fell within 50\,GHz of the transition frequency of interest (limited by photodetector speeds). Although these comb generators could not measure $f_0$ directly, $f_0$ could be constrained by locking an optical mode to the output from the last multiplication stage of a frequency chain, and by subsequent stabilization of the mode spacing directly to a microwave reference, see Fig. \ref{fig:comb_freqdomain} (This stabilization scheme yields a performance similar to the blue trace in Fig. \ref{fig:referencing} with the exception that the optical output from the frequency chain had a noise at least 100 times higher than that of $\nu_{\mathrm{opt2}}$ in Fig. \ref{fig:referencing}.)

Where comb generators required referencing in the optical domain, optical frequency combs, which yield direct access to $f_0$, allow for direct referencing and connection to the microwave domain. As such, an optical transition, $\nu_{opt}$, measured with an OFC can be expressed entirely in terms of microwave frequencies via the comb equation such that

\begin{equation}
    \nu_{\mathrm{opt}}=\nu_{\mathrm{N}} + \Delta f= N \cdot f_{\mathrm{rep}}+f_0+\Delta f.   
    \label{eq:nuopt}
\end{equation}

\noindent To use Eq. \ref{eq:nuopt}, the value of the nearest comb mode number can be determined using an optical wavemeter to loosely measure the unknown optical frequency, such that $N=$ Nearest Integer $[\nu_{\mathrm{opt(wavemeter)}}/f_{\mathrm{r}}]$. Wavemeters can yield frequency resolutions of parts in 10$^7$, or approximately 50 MHz on an optical frequency. To ensure there is no ambiguity in the mode number, OFCs with a repetition rate of > 150 MHz are typical used in atomic clock measurements.

\vspace{3mm}
\textbf{Optical clock comparisons}
A fascinating detail in optical metrology is the amazing precision with which small optical difference frequencies, such as $\Delta f$ in Eq. \ref{eq:nuopt}, can be measured against, or stabilized to, a microwave references (stabilization of an optical mode of the OFC to an optical reference is achieved by comparing and locking the small difference frequency, $\Delta f < f_{\mathrm{r}}$, to a microwave reference). Because errors on the microwave reference are additive to the optical frequency, for $\Delta f$ = 100 MHz, a $10^{-12}$ fractional error from a microwave reference yields a frequency error of 100\,$\upmu$Hz. This 100\,$\upmu$Hz error on a 300\,THz optical carrier only contributes fractionally at 3 parts in $10^{19} = 10^{-4}/3 \times 10^{14}$. To put this into perspective, using a ruler, one can split a centimeter marker by about a factor of 20, while difference frequency measurement can split 1\,GHz-spaced OFC optical modes by parts in $10^{14}$. This measurement capability is what allows optical frequency combs to measure, with no additional noise contribution, the exquisitely fine transitions of optical atomic clocks.

The measurement of absolute optical frequencies is naturally limited by the lower stability microwave references that measure them. Otherwise said, the measurement of an optical clock in terms of the Hertz is limited by the current definition of the Hertz at parts in $10^{16}$. This stability limitation has the additional drawback that averaging periods of up to a month are required to reach the maximum fractional accuracy near 1 part in 10$^{16}$ of the microwave references. To circumvent this limitation optical clocks can achieve relative measurements against one another via optical synthesis with an OFC. Because optical clocks have short-term resolutions 10 to 100 times higher than their microwave counterparts, a relative uncertainty of $10^{-16}$ can be achieved in a matter of seconds \supercite{nemitzFrequencyRatioYb2016}.

These relative clock comparisons are generally reported in the form of frequency ratios, $R = \nu_{\mathrm{opt2}}/\nu_{\mathrm{opt1}}$, which have both practical and fundamental applications to clock comparisons. Via a quick ratio measurement (See Fig. \ref{fig:comb_freqdomain}), the absolute frequency of an unknown optical clock, $\nu_{\mathrm{opt2}}(Cs)$, can be measured against a known one, $\nu_{\mathrm{opt1}}(Cs)$ as follows: 

\begin{equation}
    \nu_{\mathrm{opt2}}(Cs)= R_{2,1}\cdot \nu_{\mathrm{opt1}}(Cs).
    \label{eq:ratio}
\end{equation}
 
Here $\nu_{\mathrm{opt1}}(Cs)$ and $\nu_{\mathrm{opt2}}(Cs)$ are fractionally limited by the definition of the Hertz, but because the ratio is unit-less, $R_{\mathrm{2,1}}$ is independent of this $10^{-16}$ limit. As a result, frequency ratio uncertainties are limited by the optical transition frequencies alone. Due to the fact that optical clock transition frequencies can be controlled at levels exceeding 1 part in 10$^{-18}$ \supercite{brewer27AlQuantumlogicClock2019}, frequency ratios permit the highest precision physical measurements to date.  Additionally, because the atomic transition frequency and its dynamics are governed by the physical laws and the universal fundamental constants, atomic clock comparisons enable platforms for exquisite tests of fundamental physics \supercite{safronovaSearchNewPhysics2018}. For instance, by looking for variations in the historical time records of clock ratios, ratio measurements have placed the highest constraints on violations of special relativity \supercite{delvaTestSpecialRelativity2017, mcgrewAtomicClockPerformance2018} and searches into time variation of fundamental constants \supercite{dzubaAtomicOpticalClocks2000,bizeTestingStabilityFundamental2003, rosenbandFrequencyRatioHg2008}. Atomic clock comparisons have also been used to search for ultralight dark matter particles \supercite{heesSearchingOscillatingMassive2016} that are theorized to induce time-dependent changes in the value of the fine structure constant, and hence in the transition frequencies of some clock species. Beyond tests of fundamental physics, atomic clocks currently exhibit sensitivity to time dilation induced by the earth’s gravitational potential at centimeter relative height levels \supercite{chouFrequencyComparisonTwo2010}, impacting the relative timing between clocks at 1 part in  $10^{-18}$. It has been suggested that as clock accuracies improve, this sensitivity could be exploited to enable higher precision relativistic mapping of the geodesic potential \supercite{mehlstaublerAtomicClocksGeodesy2018}.

\subsubsection{Timing, synchronization and atomic clock networks}
\label{sec:clocknet}

Many of the applications mentioned in the previous section could be extended and benefit from global networking of optical atomic clocks \supercite{delvaTestSpecialRelativity2017}. For instance, contrasting the values of published atomic clock ratios from different national metrology laboratories currently represents the only means by which clock networks on separate continents can be compared. Additionally, the primary motivator for the development of optical clock networks is the realization of an optical SI second. However, this realization will require the networking of an international ensemble of atomic clocks for a distributed and democratic realization of optical-universal coordinated time. While microwave clocks are currently linked globally by Global Positioning Systems (GPS) and 2-way satellite time transfer, the stability of these systems can only achieve 10$^{-16}$ after one month of averaging, which is insufficient to support the improved timing capabilities of the best atomic clocks.

Aside from clock-based applications, the dissemination of ultra-high-stability timing signals is also a necessity in the context of large-scale science facilities for remote synchronization of physical events and data for high resolution measurements. As a consequence, optical fiber frequency transfer was developed to facilitate both higher stability optical clock comparisons \supercite{foremanCoherentOpticalPhase2007a} and higher resolution timing distribution and synchronization \supercite{kimDriftfreeFemtosecondTiming2008}. In the context of free-electron laser facilities, the distribution of timing signals from OFCs have enabled all-optical synchronization of remotely located lasers, accelerators, RF electronics and ultrafast X-ray experiments to within 30-femtoseconds facility-wide \supercite{schulzFemtosecondAllopticalSynchronization2015}. Additionally, over the past 15 years, fiber networks in Europe have been expanded to enable inter-city and inter-national clock comparisons of primary frequency standards at parts in 10$^{16}$ and optical clocks at the mid- 10$^{-17}$ level \supercite{lisdatClockNetworkGeodesy2016,predehl920KilometerOpticalFiber2012,guenaFirstInternationalComparison2017}. 

\begin{figure}[h!tb] 
    \centering
    \includegraphics[width=0.6\linewidth]{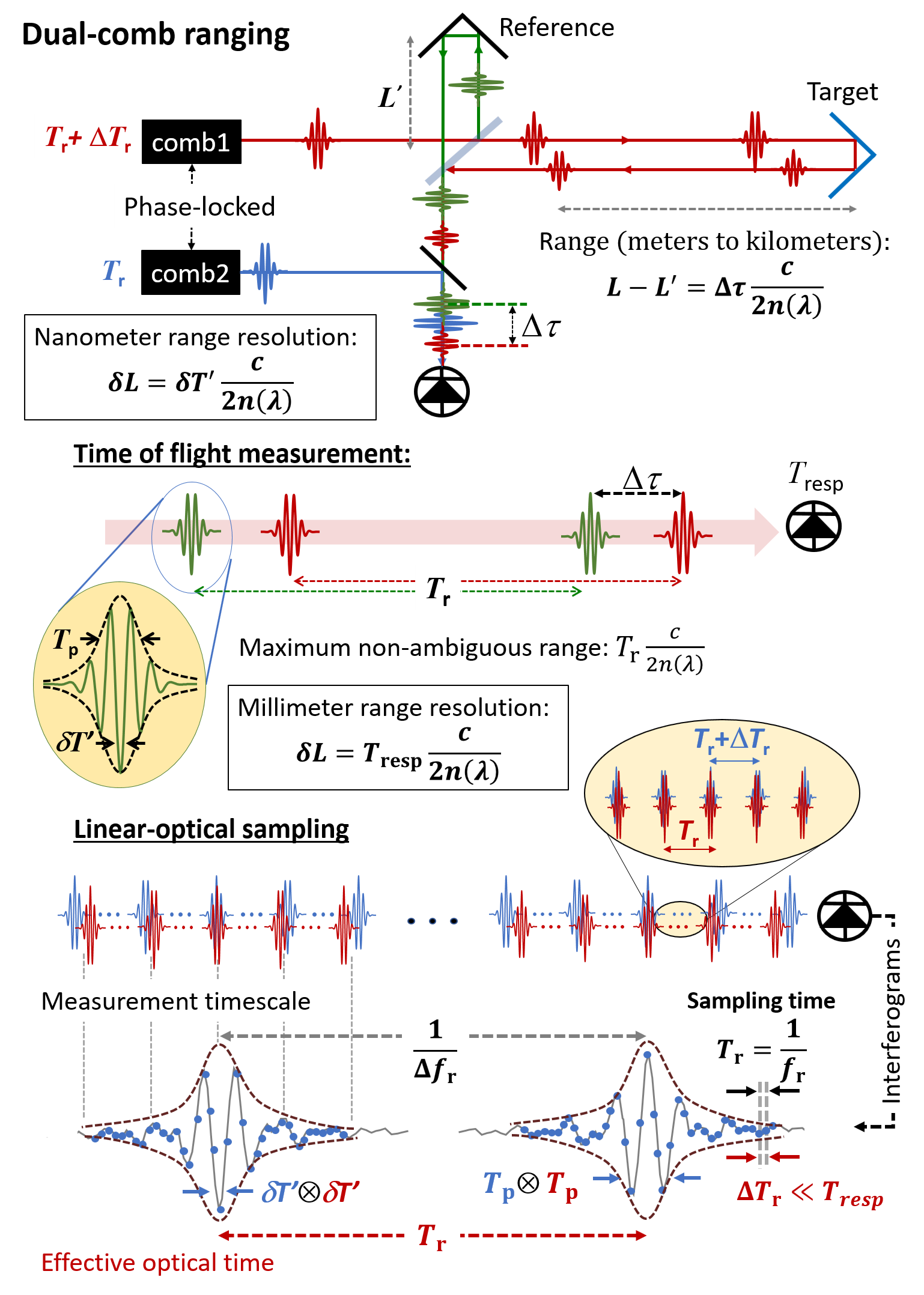}
    \caption{Dual-comb ranging and linear optical sampling. In direct time of flight ranging with a single comb, the range is determined by measuring the delay between pulses traveling to a known reference target (green pulses) and a target (red pulses). Ideally, the distance resolution, $\delta L$, should be limited by the width of the optical pulses, $T_{\mathrm{p}}$. However, when the optical pulses are detected directly, it is the much slower response time of the photodetector, $T_{\mathrm{resp}} \gg T_\mathrm{p}$, that limits the distance/timing resolution. By employing a second comb, linear optical sampling (LOS) is used to circumvent the photodetector response limit. LOS creates an interferogram between two OFCs with slightly offset repetition rates, $\Delta f_\mathrm{r}$. This optical cross-correlation between the blue and red pulses yields information about the relative optical pulse envelopes, $T_\mathrm{p}$, and the optical carriers (interferometric fringes), $\delta T'$ using microwave electronics and megahertz sampling times. The second comb for LOS improves the coarse distance resolution of direct time-of-flight measurements from the millimeter- to nanometer scale. Linear optical sampling also enables the measurement of high-resolution time/frequency information. In two-way optical time/frequency transfer, interferograms from LOS are collected at remote sites to compare, syntonize, and synchronize remote clocks, yielding time/frequency measurement accuracy and precision of parts 10$^{-17}$.}
    \label{fig:ranging}
\end{figure}

Extension of clock comparisons over long-haul fibers as a means to connect clocks from North America and Asia to Europe is unlikely due to the cost and technical difficulty to upgrade the existing infrastructure. Two-way optical time/frequency transfer (TWOTFT) across open-air paths using OFCs has been under development since 2013 \supercite{giorgettaOpticalTwowayTime2013}. These optical links allow for the comparison and synchronization of spatially separated clocks at locations where fiber networks are not readily available. Time/frequency communication is achieved by comparing counter-propagating optical pulse trains from OFCs stabilized to clocks at remote locations. Linear optical sampling (LOS), whereby each local pulse train at one site samples the incoming pulse train from the other remote site, produces interferograms that yield the relative OFC timing and frequency information (see also Fig. \ref{fig:ranging}). This permits to measure the OFC pulse arrival times from the remote site with femtosecond-level resolution using megahertz bandwidth electronics. These arrival times from both sites are used in the two-way transfer to either discern differences in clock rates, or optical path lengths fluctuations caused among others by platform motion and air turbulence. Because the optical path length fluctuations are reciprocal and thus equally sampled by the counter-propagating OFC pulse trains, comparison of optical pulse arrival times from the two sites can be used to obtain the clock rate difference. Consequently, TWOTFT enables the measurement of clock ratios, the frequency equalization of two clocks (syntonization), as well as the absolute elimination of time offset between two clocks (synchronization)\supercite{bergeronTightRealtimeSynchronization2016} with minimal residual time/frequency errors.

%**As seen in Fig. \ref{fig:ranging}, both site $A$ and $B$ probe the  arrival time $T_\mathrm{A \rightarrow B}$ respectively $T_\mathrm{B \rightarrow A}$ of the pulses from the remote site using linear optical sampling, thus generating interferograms. Because both pulse trains A$\rightarrow$B and B$\rightarrow$A sample the same air path simultaneously, timing drifts due to atmospheric turbulence and platform motion cancel out when calculating the difference of interferogram time stamps $T_\mathrm{A \rightarrow B} - T_\mathrm{B \rightarrow A}$, leaving the timing difference $\Delta T_\mathrm{AB}$ between both clocks. This allows to measure clock ratios, equalize the frequency of two clocks (syntonization) or eliminate the absolute time offset between two clocks (synchronization)\supercite{bergeronTightRealtimeSynchronization2016}.**

For example, TWOTFT links have been implemented up to several kilometers and have demonstrated timing errors as low as 7\,as after 1\,s of averaging  \supercite{sinclairComparingOpticalOscillators2018}. While all OFC based TWOTFT demonstrations to date used kilometer-scale terrestrial links, it has been inferred \supercite{swannMeasurementImpactTurbulence2019} that a TWOTFT ground-to-satellite link should only increase the turbulence-induced timing noise to a few femtoseconds. Finally, the low duty cycle of the optical pulses make the link significantly less sensitive to interruptions (for times less than 1/$f_{\mathrm{r}}$, usually a few nanoseconds) than those based on continuous wave signals. 

\subsection{Ultra low-noise microwave generation}
\label{sec:mwave}

As discussed in Section \ref{sec:OFCstab}, OFCs stabilized to high stability optical references can enable the derivation of microwave signals with stabilities better than $10^{-15}$, yielding greater than a 100 times improvement over what can be achieved with the best room-temperature electronic oscillators \supercite{diddamsDesignControlFemtosecond2003}. In principle, the generation of low-noise microwave is relatively straight forward. As shown in Fig. \ref{fig:comb_timedomain}, the phase-stabilized optical pulses from an OFC can be converted to a stable electronic pulse train via direct detection with high-speed photodetectors. Subsequent electronic filtering isolates a single harmonic of $f_\mathrm{r}$ within the bandwidth of the photodetector producing a sinusoidal signal. Signals derived in this manner can be utilized as microwave frequency representation of optical atomic clocks ("ticks") to facilitate their characterization against primary microwave standard and for the generation of microwave timing signals.

An additional benefit of optical frequency division (OFD), as described in Eq. \ref{eq:locking1}, is that OFD also divides the noise of the optical reference to the microwave domain, yielding high-spectral purity electronic signals. For example, optical frequency division of high-stability optical cavities have been used to generate 10\,GHz signals with ultra-low phase noise of -100\,dBc/Hz at 1\,Hz (stepping 1\,Hz off the microwave carrier, the spectral noise density falls 10 orders of magnitude below the carrier)\supercite{milloUltralownoiseMicrowaveExtraction2009,fortierGenerationUltrastableMicrowaves2011}. Ultra-low noise electronic signals have natural applications to military, communications and test and measurement applications. For example, in X-band RADAR \supercite{capmanyMicrowavePhotonicsCombines2007} signals with low close-to-carrier noise can facilitate the detection of slow moving objects (with Doppler shifts below 1\,kHz), and the measurement of weak return signals from low-cross section objects. Additionally, low noise on local oscillators can enable shorter averaging periods for quicker acquisition of targets and frequency-hopped communications signals.

To put the potential for low-noise into perspective, self-referenced optical frequency combs can produce optical pulses with > 200\,dB of optical dynamic range \supercite{benedickOpticalFlywheelsAttosecond2012} and sub-femtosecond pulse-to-pulse timing jitter. Ultimately, the biggest challenge in low-noise microwave generation is preservation of these exquisite optical pulse characteristics in their conversion to the microwave domain. Aside from the massive pulse distortion that results due to limited detector response time and power handling, the detection of high-energy optical pulses further exacerbate detector nonlinearities by creating short bursts of extremely high-density photo-carriers in the detector active region. Normally, this results in charge-screening of the applied bias voltage, which reduces detector speed and yields strong conversion of optical amplitude noise to phase noise on the microwave pulse train \supercite{taylorCharacterizationPowertoPhaseConversion2011}. Consequently, to attain the highest-spectral purity microwave signals, low-pulse energy, higher-repetition rate OFC sources ($f_{\mathrm{r}}$ > 1\,GHz) are ideal \supercite{habouchaOpticalfiberPulseRate2011}. These distortion and saturation effects can be mitigated by detectors with heterostructures designed to improve charge transport speeds and that pre-distort the internal potential for high-linearity at high-photocurrents \supercite{belingHighpowerHighlinearityPhotodiodes2016}. Impressively, such photodetectors have demonstrated preservation of the optical pulse-to-pulse timing at parts in $10^{17}$ \supercite{baynesAttosecondTimingOpticaltoelectrical2015} and have enabled microwave signal dynamic ranges as high as 180\,dBc \supercite{quinlanExploitingShotNoise2013,xiePhotonicMicrowaveSignals2017}. Finally, a significant side benefit of ultra-short optical pulses (< 1 picosecond) is that coherence of the optical modes results in modified shot noise statistics, or cyclo-stationary noise, which permits microwave phase noise floors significantly lower than what is possible with CW lasers\supercite{quinlanExploitingShotNoise2013}.

\subsection{Calibration of astronomical spectrographs}
\label{sec:astroOFC}

Optical frequency combs were first proposed for improved frequency calibration of astronomical spectrographs in 2007 \supercite{murphyHighprecisionWavelengthCalibration2007}. These spectrographs measure Doppler shifts of stellar spectra to determine the radial velocities of celestial bodies, as well as the composition of solar atmospheres. Higher precision measurements, required to discern the drift in these Doppler shift on the order of $\sim$ 1\,cm/s, can be used to assess the rate of expansion of the universe, as well as detection of the periodic wobble in stellar velocity due to the influence of an orbiting exo-planet. These small Doppler shifts are inaccessible with conventionally calibrated spectrographs due to environmental instability of the spectrographs, and due to the limited spectral coverage and stability of traditional calibration sources (wavelength references).  The observation of the physical locations of the exactly equidistant modes from an OFC at the spectrograph's imaging plane allows for calibration of instrument drift and improved long-term frequency accuracy of the spectrograph. 

The application specifications require that these astro-OFCs overcome some technical challenges, namely high-mode spacings near 30\,GHz for matching to the spectrographs resolving power, as well as very broad and spectrally-flat coverage ($\Delta P$ < 3\,dB) from 380\,nm to 2.4\,$\upmu$m \supercite{mccrackenDecadeAstrocombsRecent2017} for uniform illumination of the spectrograph, and frequency stability at a level of $3\times 10^{-11}$ corresponding to 1\,cm/s radial Doppler drift.
As such, different architectures have being developed using mode-filtered fiber \supercite{steinmetzLaserFrequencyCombs2008} and Ti:Sapphire based OFCs, and more recently electro-optic comb generators \supercite{metcalfStellarSpectroscopyNearinfrared2019}. Optical frequency combs began deployment to spectrographs in 2008 \supercite{steinmetzLaserFrequencyCombs2008}, and can be currently found in at least seven telescopes around the world \supercite{fischerStateFieldExtreme2016,mccrackenDecadeAstrocombsRecent2017}. Of these, the high precision OFC calibrations were performed to 2.5\,cm/s at HARPS/FOCES \supercite{probstRelativeStabilityTwo2016} and to $< 10$\,cm/s at HARPS-N \supercite{glendayOperationBroadbandVisiblewavelength2015} and the Hobby–Eberly Telescope \supercite{metcalfStellarSpectroscopyNearinfrared2019}.

\subsection{Optical frequency combs beyond the laboratory}
\label{sec:outoflab}

\subsubsection{Distance measurements and laser ranging}
\label{sec:distance}

The application of optical frequency combs to LIDAR (light detection and ranging) was first demonstrated in 2000 \supercite{minoshimaHighaccuracyMeasurement240m2000} and enables a number of advantages over traditional sources. As seen in Fig. \ref{fig:ranging}, two techniques can be used to discern macroscopic distances: the more coarse direct time-of-flight measurement, and the more fine linear optical sampling (LOS) measurement. As seen in Fig. \ref{fig:ranging}, the maximum distance resolution in direct time of flight measurements is limited by the response time of the photodetector, which is much slower than the optical pulse envelope width, or $T_{\mathrm{resp}} \ll T_{\mathrm{p}}$. Ranging measurements, enabled by dual-comb techniques (see also section \ref{sec:CombSpectroscopy}) or balanced optical cross-correlation in nonlinear crystals, can bypass the limitations of $T_{\mathrm{resp}}$ via optical down-sampling so that the optical carrier can be detected on low-speed photodetectors (Fig. \ref{fig:ranging}). This permits distance measurements with resolutions limited by the width of the convolution of the two OFC pulse envelopes, $T_{\mathrm{p}} < 100$ s, significantly shorter than the $\sim$1\,ns pulses from traditional LIDAR systems. These gains in resolution and precision can be further improved by detecting the optical carrier under the pulse envelope. Interferometric measurements of this kind have demonstrated 10\,nm resolution for averaging times of $\sim$60\,ms, and nanometer-level precision with longer averaging over second timescales \supercite{coddingtonRapidPreciseAbsolute2009, minoshimaHighaccuracySelfcorrectionRefractive2011}.
%In less controlled environments, because the OFC spectrum offers a large optical bandwidth, the index of refraction of air can be measured by knowing the relative phase differences of various wavelength components, and hence correct for uncertainties that result due to air turbulence\supercite{minoshimaHighaccuracySelfcorrectionRefractive2011, jangCompensationRefractiveIndex2017}. 
Finally, the combination of short pulses, and a phase-stabilized optical spectrum has allowed for demonstrations of comb-based distance measurements at kilohertz measurement rates and sub-micrometer ranging precision, while simultaneously enabling unambiguous range measurement to targets at meter to kilometer distances \supercite{jangDistanceMeasurementsUsing2018,coddingtonRapidPreciseAbsolute2009,zhuDualCombRanging2018}.

The largest drawback of using optical frequency combs directly for ranging is their high cost and high system complexity. Consequently, recent work has pushed toward the development of smaller and simpler systems based on electro-optic and micro-resonator comb systems \supercite{wuLongDistanceMeasurement2017, trochaUltrafastOpticalRanging2018}, as well as on photonically-integrated circuits \supercite{weimannSiliconPhotonicIntegrated2017}. A challenge for deployment of OFC-based systems to the field, similar to that in dual-comb spectroscopy (\ref{sec:CombSpectroscopy}), is that the low-power per optical mode requires a retro-reflecting mirror for sufficient return light powers at distances beyond $\sim$1\,m.

\subsubsection{Direct comb-based spectroscopy}
\label{sec:CombSpectroscopy}
\begin{figure}[h!tb] 
    \centering
    \includegraphics[width=0.9\linewidth]{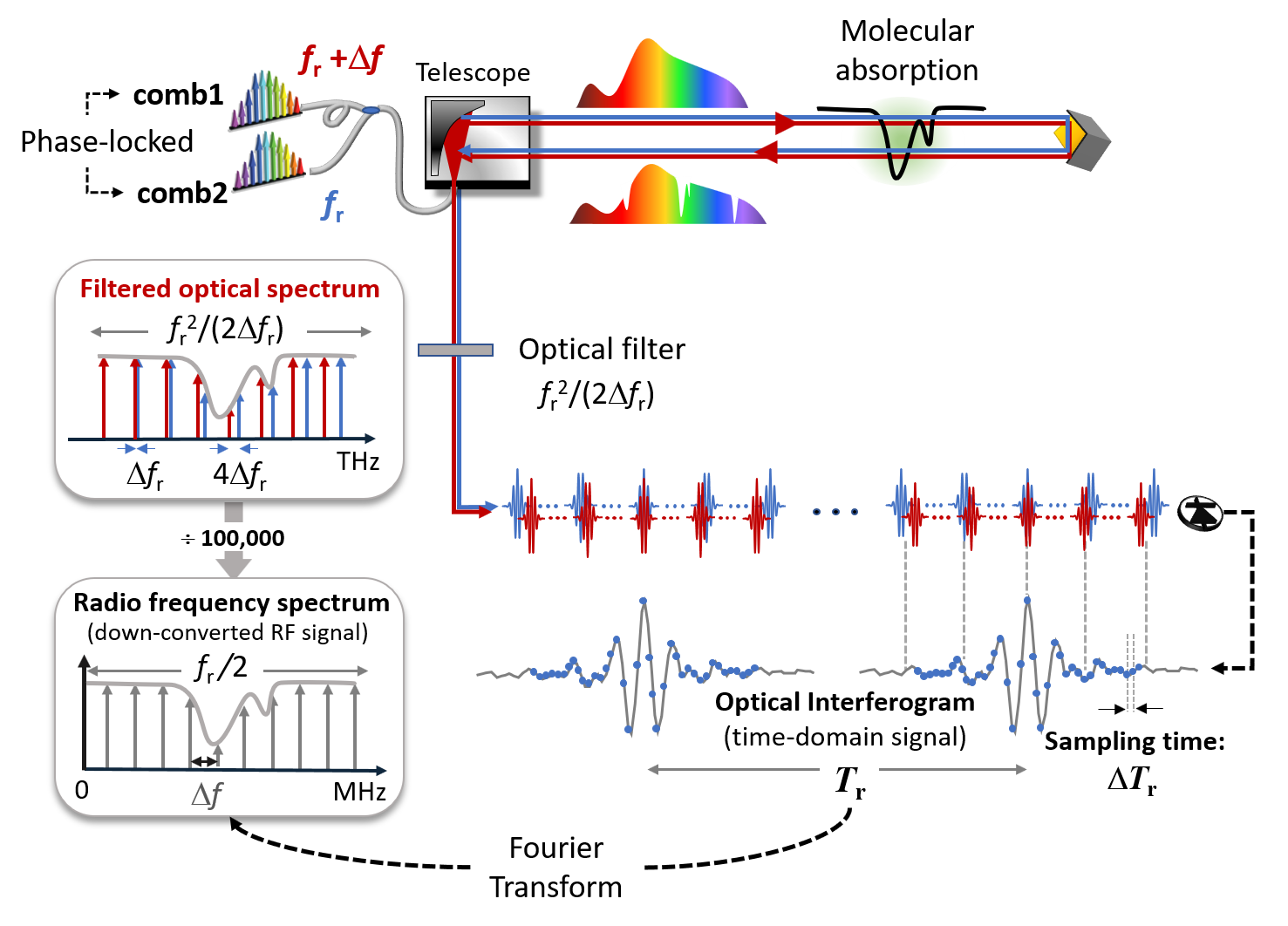}
    \caption{Dual-comb spectroscopy. The output from spatially overlapped optical fiber frequency combs with repetition rates $f_r\sim$100\,MHz are phase stabilized to one another with repetition rates slightly differing by $\Delta f_{\mathrm{r}}\sim$1-10\,kHz. The pulse train from one or both OFCs traverse an atmospheric path or a gas cell, thus encoding the spectroscopic information as amplitude and phase on the optical pulse trains and comb modes. Linear optical sampling is used to produce optical interferograms to extract optical spectroscopic information. Due to the slight difference in repetition rate $\Delta f_{\mathrm{r}}$ between the OFCs, the pulse trains slowly raster through one another, yielding the interferogram, which is a slowly time varying interference signal. When photo-detected, this signal repeats with a periodicity of $1/\Delta f_\mathrm{r}$ (see also Fig. \ref{fig:ranging}). Notably, assuming invariant pulses over time scales of $1/\Delta f_r$, this slow advancement of comb2 pulses by $\Delta T_{\mathrm{r}}=1/f_\mathrm{r}-1/(f_\mathrm{r}+\Delta f_\mathrm{r})\approx$ 100\,fs relative to comb1 pulses for each comb pair effectively downsamples the optical signal by $f_{\mathrm{r}}/\Delta f_{\mathrm{r}}$ $\sim$100,000, yielding femtosecond optical fringe information to megahertz bandwidths that can be processed with off-the-shelf electronics. Fourier Transform of the optical interferogram yields a microwave comb RF spectrum with mode spacing $\Delta f_{\mathrm{r}}$. All spectral information from the optical domain is mapped to an RF bandwidth $\leq f_{\mathrm{r}}/2$, corresponding to an optical span of $f_{\mathrm{r}}^2/(2\Delta f_{\mathrm{r}})$. Spectral filtering is usually applied to ensure non-ambiguous detection over this spectral span.}
    \label{fig:DCS}
\end{figure}

Frequency combs are attractive sources for spectroscopy because they offer: 1) broad spectral bandwidth, good for detection of multiple molecular species, 2) high spatial coherence, which allows for longer interrogation paths and higher sensitivity, and 3) high frequency resolution and accuracy when measured on a mode by mode basis. As a result, spectroscopic measurement with OFCs has been explored over vast portions of the electro-magnetic spectrum, extending from the terahertz to the ultraviolet. A concerted effort has also been devoted for coverage in the mid-IR region, to access stronger molecular cross-sections 
\supercite{klockeSingleShotSubmicrosecondMidinfrared2018,muravievMassivelyParallelSensing2018,schliesserMidinfraredFrequencyCombs2012,ycasHighcoherenceMidinfraredDualcomb2018, kowligyInfraredElectricField2019}, and OFC based spectroscopy has also been covered extensively in various review articles (see for example \supercite{cosselGasphaseBroadbandSpectroscopy2017,weichmanBroadbandMolecularSpectroscopy2019,coddingtonDualcombSpectroscopy2016,adlerCavityenhancedDirectFrequency2010}.) More specifically, direct spectroscopy with OFCs has been applied to study ultra-cold molecules \supercite{changalaRovibrationalQuantumState2019}, human breath analysis \supercite{thorpeBroadbandCavityRingdown2006}, time-resolved spectroscopy \supercite{ideguchiCoherentRamanSpectroimaging2013,lomsadzeTricombSpectroscopy2018a}, and high-precision molecular spectroscopy \supercite{longMultiplexedSubDopplerSpectroscopy2016a,nishiyamaPreciseHighlysensitiveDopplerfree2018}.

While multiple comb-based measurement techniques have been explored for molecular spectroscopy, such as comb-assisted Fourier transform infrared spectroscopy (FTIR) \supercite{maslowskiSurpassingPathlimitedResolution2016} and virtually-imaged-phased-array (VIPA) \supercite{diddamsMolecularFingerprintingResolved2007}, the most notable has been dual-comb spectroscopy (DCS) \supercite{coddingtonDualcombSpectroscopy2016}. Dual-comb spectroscopy, which was first demonstrated in 2004, \supercite{keilmannTimedomainMidinfraredFrequencycomb2004,schliesserFrequencycombInfraredSpectrometer2005} using linear-optical sampling (LOS) of two combs with slightly offset repetition rates to down-convert 10’s of terahertz of optical bandwidth to 100's of megahertz in the RF domain (see Fig. \ref{fig:DCS}). This detection scheme preserves the stability and accuracy of the comb, as well as the amplitude and phase information encoded on the individual modes of the comb by the spectroscopic sample. Consequently, the detected RF spectrum can be used to reconstruct the optical molecular absorption spectra.

Direct molecular spectroscopy with DCS enables significant benefits as compared to Fourier transform infrared spectroscopy. Perhaps the biggest advantage is that DCS does not require any moving parts, thus enabling much faster acquisition times. Stabilization of the OFC spectrum also permits much higher resolutions, as well as SI-traceability for higher accuracy. Finally, because there are no complicated diffraction and imaging optics, measurements do not suffer instrument lineshapes. While there are many benefits to DCS, challenges arise for deployment to the field. The low power per optical mode requires a retro-reflecting mirror to obtain strong enough signals strengths, and spectral magnitude fluctuations pose challenges, especially when interrogating broad molecular transitions, or while probing a cluttered molecular spectrum over the atmosphere.

\section{Perspective and future outlook}

While precision measurements may seem like a boutique application primarily reserved for metrology laboratories, this is the area where combs have seen the most commercial success, in part because large-scale laboratory experiments are willing to pay the high price for performance. The adoption of optical frequency comb products for replacement of traditional LIDAR and FTIR systems, however, has yet to find commercial traction. Combs, which offer better precision, resolution and faster acquisition times for both applications, yield a cost and complexity that is still too high to outweigh its benefits. As a result, this cost-benefit trade-off has helped to fuel the development of compact and lower SWAP spectroscopic sources \supercite{suhMicroresonatorSolitonDualcomb2016, duttOnchipDualcombSource2018,yuGasPhaseMicroresonatorBasedComb2018,sterczewskiTerahertzHyperspectralImaging2019,linkDualcombSpectroscopyWater2017,millotFrequencyagileDualcombSpectroscopy2016,klockeSingleShotSubmicrosecondMidinfrared2018}.

In precision metrology, the replacement of frequency multiplier chains by OFCs enabled a vast simplification to precision optical measurement, helping the development of new clock species, as well as facilitating continued improvement in optical clock accuracy. In the past 20 years, optical atomic clock accuracies have improved by five orders of magnitude, and now demonstrate performance 100 times better than state-of-the art primary $^{133}$Cs microwave standards. This rapid progress has prompted discussions about, and the creation of a road-map toward the redefinition of the SI second to optical atomic time \supercite{riehleCIPMListRecommended2018}. In the context of future optical timescales, optical frequency combs will provide the clockwork for the derivation and dissemination of highly-precise and accurate microwave and optical timing signals. 

In the near future, optical frequency combs, integrated with transportable optical clocks, will be important for supporting work toward optical redefinition of the second by facilitating clock comparisons between remotely located metrology labs \supercite{grottiGeodesyMetrologyTransportable2018}. Aside from global clock comparisons, comb-based inter-continental time/frequency-transfer between laboratory based clocks and transportable clocks could help to enable high precision mapping of the geoid, and help to facilitate tests of fundamental physics that can leverage very long baselines. To date, while high-precision clock comparisons have yet to uncover dark matter signatures or observe temporal variations of fundamental constants, these tests have been used to verify our understanding of current physical models at parts in 10$^{17}$. In the future, ranging and time-transfer with space-borne combs \supercite{leziusSpaceborneFrequencyComb2016,leeTestingFemtosecondPulse2014} might be used for ground to satellite clock synchronization for improved timing in GNSS and communications systems, and sensing with combs could be used for atmospheric spectroscopy in broadband occultation. Finally, tests of fundamental physics beyond parts in $10^{19}$ might one day be enabled by space-borne systems benefiting from operation in a low-vibration environment and outside the earth's gravitational potential, as well as from non-classical statistics \supercite{giovannettiAdvancesQuantumMetrology2011} enabled by quantum combs \supercite{leeFrequencyCombSinglephoton2018}...

\vspace{5mm}
\textbf{Acknowledgements}. We acknowledge fruitful discussion with William Loh, Fabrizio Giorgetta, Gabe Ycas, Ian Coddington, William Swann, and Florian Adler. We would also like to thank Fabrizio Giorgetta, Holly Leopardi, Megan Kelleher, Nick Nardelli, Frank Quinlan, Tanja Cuk, Ladan Arissian, and Henry Timmers for thorough reading of the manuscript.

\textbf{Data availability}
No data set were generated or analyzed during the current study.
% \bibliography{CombReview_bibtex}
% \bibliographystyle{naturemag}
\printbibliography

\textbf{Author Contributions}
T.M.F. and E.B. equally contributed to the ideas, organization and writing of this review article.

\end{document}